\documentclass[aps,prd,nofootinbib,twocolumn,superscriptaddress,preprintnumbers,balancelastpage,longbibliography]{revtex4-1}

\usepackage{bm}
\usepackage{appendix}
\usepackage[utf8]{inputenc}
\usepackage{amsmath,amssymb,mathtools,bm}
\usepackage{graphicx, color, hepunits}
\usepackage[dvipsnames]{xcolor}
\usepackage{float}
\usepackage{multirow}
\usepackage{tikz-feynman} 
 \usepackage{hyperref}
\hypersetup{
    colorlinks=true,       
    linkcolor=blue,        
    citecolor=blue,        
    filecolor=magenta,     
    urlcolor=blue          
}
\usepackage[english]{babel}
\let\vec\mathbf
\usepackage{tensor}
\usepackage{orcidlink}

\renewcommand{\eqref}[1]{Eq.~(\ref{#1})}

\newcommand\Tstrut{\rule{0pt}{2.5ex}}         
\newcommand\Bstrut{\rule[-1.3ex]{0pt}{0pt}}

\newcommand{\vecv}{\mathbf{v}}
\newcommand{\vobs}{\mathbf{v}_{\rm obs}}
\newcommand{\frms}{f_{\rm RMS}}

\newcommand{\vecb}[1]{\ensuremath{\boldsymbol{#1}}}
\DeclareMathOperator\erf{erf}

\begin{document}

\title{Statistics of Daily Modulation in Dark Matter Direct Detection Experiments}

\author{Carlos Blanco~\orcidlink{0000-0001-8971-834X}}
\affiliation{Institute for Gravitation and the Cosmos, The Pennsylvania State University, University Park, PA 16802, USA}
\affiliation{Institute for Computational and Data Sciences, Penn State University, University Park, PA 16802, USA}
\affiliation{Department of Physics, Princeton University, Princeton, NJ 08544, USA}

\author{Joshua W. Foster~\orcidlink{0000-0002-7399-2608}}
\affiliation{Department of Physics, University of Wisconsin-Madison, Madison, WI 53706, U.S.A}

\author{Yonatan Kahn~\orcidlink{0000-0002-9379-1838}}
\affiliation{Department of Physics, University of Toronto, Toronto, ON M5S 1A7, Canada}

\author{Benjamin Lillard~\orcidlink{0000-0001-8496-4808}}
\affiliation{{Department of Physics, University of Oregon, Eugene, OR, 97403, USA}}
\affiliation{Institute for Gravitation and the Cosmos, The Pennsylvania State University, University Park, PA 16802, USA}

\date{\today}

\begin{abstract}
The time-dependent modulation of the event rate in dark matter direct detection experiments, arising from the motion of the Earth with respect to the Galactic rest frame, is a distinctive signature whose observation is crucial for claiming a discovery of dark matter. While annual modulation has been well studied for decades, daily modulation due to the Earth's rotation has attracted increased attention recently due to the identification of anisotropic solid-state detector materials that yield a direction-dependent scattering rate without sacrificing the overall rate. We perform a statistical analysis of daily modulation in dark matter scattering experiments, with the goal of maximizing the statistical significance of a modulating signal in the presence of an unknown background rate, which may be either flat (non-modulating), or modulating over a 24-hour period with a known or unknown phase. In the background-dominated regime, we find that the discovery significance scales as $\frms\sqrt{T}$, where $T$ is the total exposure time and $\frms$ is the root-mean-square modulation amplitude; in particular, the significance continues to improve with exposure rather than saturating due to systematic uncertainties in the background rate. Using anisotropic trans-stilbene detectors for sub-GeV dark matter as a benchmark example, we provide prescriptions for optimizing the significance for a given total detector mass and location. In an example analysis using three detectors, optimizing the detector orientations can reduce the required exposure by a factor of $\sim 5$ for a desired discovery or exclusion significance, even after profiling over an unknown modulating background phase.
\end{abstract}
\maketitle

\tableofcontents

\section{Introduction}

While the particle nature of dark matter (DM) remains unknown~\cite{Akerib:2022ort, PICO:2019vsc, DarkSide-50:2022qzh,  LZ:2024zvo, XENON:2025vwd, PandaX:2025rrz, SENSEI:2023zdf,DAMIC-M:2025luv}, \textit{N}-body simulations of its gravitational interactions yield a fairly consistent description of its velocity distribution~\cite{Folsom:2025lly,Wojtak05,Hansen06,Vogelsberger08,Vogelsberger09,Kuhlen10,Green10,Vogelsberger13,Ling10,Tissera10,Pillepich14,Butsky16,Kelso16,Sloane16,Bozorgnia16,Bozorgnia17,Poole-McKenzie20,Bozorgnia20,Hryczuk20,Nunez-Castineyra23,Lawrence23,Staudt24}. Under quite general assumptions, the bulk of the DM velocity distribution $f(\vecv)$ is expected to be largely isotropic in the Galactic rest frame, such that the distribution observed on Earth, $f(\vecv + \vobs(t))$, acquires a time dependence thanks to the detector velocity $\vobs(t)$ with respect to the Galactic frame:
\begin{equation}
    \vobs(t) = \vecv_{\odot} + \vecv_{\oplus}(t),
    \label{eq:vobs}
\end{equation}
where $\vecv_{\odot}$ is the Sun's velocity with respect to the Galactic center, and $\vecv_{\oplus}(t)$ is the Earth's velocity with respect to the Sun~\cite{Lewin:1995rx}. Changes in $|\vecv_{\oplus}(t)|$ over the course of a year lead to $\sim 10\%$ changes in the maximum DM speed in the lab frame, yielding an associated annual modulation in the total event rate which has been well-studied for decades~\cite{Freese:2012xd}. Due to the rotation of the Earth, the lab-frame $\vecv_{\oplus}(t)$ changes direction over the course of a sidereal day, giving rise to a daily modulation in the event rate in detectors that are sensitive to the direction of the incoming DM~\cite{Mayet:2016zxu}. 
Directional detectors for traditional WIMPs infer the incoming DM direction from the momentum of the outgoing scattered nucleus~\cite{Vahsen:2021gnb}, but the need for either a dilute medium, such as a gaseous detector~\cite{Vahsen:2020pzb}, or an energy deposit very near a material-specific threshold~\cite{Bozorgnia:2011tk,Kadribasic:2017obi} typically leads to a lower overall event rate compared to non-directional detectors. 

Over the past decade, though, there have been numerous proposals to detect sub-GeV DM with anisotropic solid-state detector materials, where the incoming DM direction affects the overall excitation rate through anisotropies in the transition matrix elements without the need to track any final-state momentum directions. For MeV--GeV DM, detector materials such as anisotropic organic scintillators~\cite{Blanco:2021hlm} and diatomic molecules~\cite{Blanco:2022pkt} can yield overall event rates within an order-1 factor of traditional isotropic detector materials such as silicon or germanium, with the added benefit of a $\sim 10\%$ daily modulation amplitude. For sub-MeV dark matter, narrow-gap~\cite{Hochberg:2017wce,Coskuner:2019odd,Abbamonte:2025guf,Griffin:2025wew,Hochberg:2025dom,Hochberg:2025rjs} or polar~\cite{Griffin:2018bjn,Coskuner:2021qxo} materials often have some degree of anisotropy, such that any experiment with sub-eV energy thresholds may naturally observe a modulating event rate. That said, novel detector materials and techniques are inevitably accompanied by new background sources, such as the persistent low-energy excess~\cite{Baxter:2025odk}; achieving the lowest threshold and thus the largest DM mass reach possible in a zero-background environment may not be possible with these new detectors. As the sub-GeV parameter space continues to be probed through detectors with larger exposures~\cite{SENSEI:2023zdf,DAMIC-M:2025luv,SuperCDMS:2025dha,Oscura:2022vmi}, it is plausible that a DM discovery may come through a modulating event rate, likely accompanied by unknown background rates which may be flat (i.e.\ time-independent) and/or modulating on a period of an Earth day rather than a sidereal day.

In this paper, we set up the statistical framework to enable DM discovery through daily modulation in the presence of unknown backgrounds. We focus on the realistic near-term scenario of a counting experiment with a known modulating signal profile, and construct a binned Poisonnian likelihood function that can account for either a small or large number of events per time bin.\footnote{In particular, we do not consider wave-like DM experiments in this work, for which the noise statistics are quite different; see~\cite{Knirck:2018knd,Foster:2020fln,Foster:2023bxl} for detailed analyses of modulation effects in wave-like DM experiments.} To make the discussion concrete, we construct realistic daily modulation curves $R(t)$ for DM--electron scattering in the trans-stilbene detectors proposed in Ref.~\cite{Blanco:2021hlm}, and we demonstrate how to build joint likelihood functions appropriate for a multiplexed detector consisting of crystals with varying signal-rate profiles, subject to both correlated and uncorrelated backgrounds in each detector. 

Our main result is that the figure of merit for the statistical significance of a modulating signal generically scales as $\frms\sqrt{T}$, where $T$ is the total exposure and $\frms$ is the root-mean-square modulation amplitude, which for a flat (time-independent) background is defined as 
\begin{equation}
    \frms \equiv \sqrt{\int_0^T \frac{dt}{T} \big( r(t) - 1 \big)^2 }.
    \label{eq:frms}
\end{equation}
for a dimensionless signal profile normalized such that $\int_0^T \frac{dt}{T} r(t) = 1$. We generalize this result to arbitrary background profiles and develop a formalism for obtaining the expected sensitivity for an experiment employing an arbitrary numbers of directional detectors, each with its own signal profile. Previous work~\cite{Geilhufe:2019ndy,Blanco:2021hlm} focused on a two-bin analysis, but we show how one may perform either a stacked binned analysis with an arbitrary number of bins, or an unbinned analysis. While other statistical treatments of daily modulation have appeared in the DM literature before~\cite{Griffin:2018bjn, Coskuner:2019odd, Coskuner:2021qxo}, our goal in this work is to provide a prescriptive formalism for performing daily modulation analyses in direct detection experiments, focusing on the case of multiple detectors in order to maximize the statistical power through distinct signal profiles for each orientation. We show through several examples that order-1 improvements in the discovery significance may be possible by optimizing the detector orientations based on priors for the relative amplitude and time variation of the dominant backgrounds. We expect these considerations may be useful for quantifying the trade-off between overall detector mass, modulation amplitude, and background sources, informing future designs of directional detectors leveraging novel anisotropic materials.

This paper is organized as follows. In Sec.~\ref{sec:SingleLikelihood}, we construct the likelihood for a single detector orientation in the presence of a flat background rate or a time-varying background rate, and derive the scaling of the significance with modulation fraction and exposure, in the regime of a large number of counts per time bin. In Sec.~\ref{sec:MultiLikelihood}, we generalize to multiple detectors (each with their own modulating signal profile), construct a joint likelihood, and use the Fisher information matrix to quantify the discovery significance. In Sec.~\ref{sec:Binning}, we study the effect of varying the time binning of the data, demonstrating through Monte Carlo that the parametric scaling of the test statistic is valid even in the narrow-bin limit when the Gaussian approximation for individual bin counts no longer holds. If a binned analysis is desired, Sec.~\ref{sec:Sidereal} presents a scheme for stacking the data on the period of a sidereal day which isolates a modulating DM signal from potential backgrounds which modulate on an Earth day. In Sec.~\ref{sec:Example}, we perform an example analysis for a multiplexed trans-stilbene detector and demonstrate the advantages of optimizing the orientations of multiple anisotropic detectors, each with their own time-dependent signal profile. We conclude in Sec.~\ref{sec:Conclusions}.

\section{Single Detector Likelihood}
\label{sec:SingleLikelihood}
We begin by developing the formalism for a likelihood-based analysis as relevant for a single detector. We assume the detector is a counting experiment, which detects events that may have been generated either by DM scattering or by mundane Standard Model (SM) processes. We further assume that the SM background rate of the experiment is not precisely known. That is, whether or not we have an estimate for the dominant sources of noise, we cannot directly measure the number of DM events by subtracting the estimated SM background. This set of assumptions is appropriate for virtually all of the state-of-the-art DM direct detection experiments, and is the motivation behind directional detectors, since the unknown background rate presents a systematic ``fog'' to discovery sensitivity for non-directional detectors~\cite{Akerib:2022ort}. 

We treat both signal and background processes as inhomogeneous Poisson processes with time-dependent rates $s(t)$ and $b(t)$. It is natural to frame any DM analysis in terms of an overall signal strength  parameter, which we denote $A$, that controls the overall scale of signal process, acting as a proxy for, \textit{e.g.}, the DM scattering cross section, the DM density, or other model-dependent factors. We will also take the rate to have no additional parameter dependence; in particular, we assume that the DM velocity distribution is fixed and not marginalized over in this analysis.

Under this parametrization, the total counts as measured by the detector are drawn from an inhomogeneous Poisson process with total rate $r(t)$ given by
\begin{equation} \label{eq:rtAsb}
    r(t| \{A, \bm{\Theta}\}) = A \, s(t) + b(t|\bm{\Theta}).
\end{equation}
where $b(t)$ is specified by additional nuisance parameters $\bm{\Theta}$. In this work, we work in the weak-signal limit such that $A \, s(t) \ll b(t)$. This is, of course, the physically-relevant regime in which evidence for a weak signal may eventually emerge above otherwise dominant backgrounds, but it also computationally important for two reasons. First, it will allow us to work at leading order in the signal. Second, in some cases an analysis will permit $A\, s(t) < 0$.  By construction, the weak-signal limit ensures that even a negative signal rate does not cause the overall rate $r(t)$ to become negative.

\subsection{Chi-square derivation}
\label{sec:chi-square}

Our hypothetical experiment measures uncorrelated events with some time resolution $\Delta t$. For a daily modulation experiment we require $\Delta t \ll 1$\,day, but the typical size of $\Delta t$ depends on the detector technology: for Skipper CCDs it may be several hours~\cite{SENSEI:2023zdf}, while for standard WIMP experiments~\cite{PICO:2019vsc, DarkSide-50:2022qzh, LZ:2024zvo, XENON:2025vwd} it is effectively instantaneous. 
In this section, we will begin each analysis by gathering the data into time bins of duration $t_\text{bin} \lesssim 10^{-1}$ day, so as long as $\Delta t \ll t_\text{bin}$ the actual value of $\Delta t$ is unimportant. 
In the limit of arbitrary precision ($\Delta t \rightarrow 0$), the $i$th event has its own timestamp $t_i$, and an experimental outcome is a list of $N$ timestamps,  $\{t_1, t_2, \ldots, t_{N}\}$. 
We sort these data into $N_{\rm bins}$ bins, and take $d_j$ to be the number of events in the $j$th bin, with 
\begin{align}
\sum_{j=1}^{N_{\rm bins}} d_j = N
\end{align}
total events. For simplicity, we take the exposure times in each bin to be equal.

Given timestamped data $\{ d(t_1), \ldots, d(t_N) \}$ or its binned histogram $\vec d \equiv \{d_1, \ldots, d_{N_{\rm bins}}\}$, our goal is to determine whether a hypothesis $\vecb\mu = \{ \mu_1, \mu_2, \ldots , \mu_{N_{\rm bins}}\}$ is a good fit to the data. Under our assumptions, the probability that the experiment would measure $d_j$ events in the $j$th bin is given by the Poisson distribution,
\begin{equation}
\mathcal L(d_j | \mu_j) = \frac{\mu_j^{d_j} e^{-\mu_j }}{ d_j!}.
\end{equation}
The likelihood function for the binned data is the Poisson likelihood,
\begin{equation}
\mathcal L(\vec d | \vecb{\mu}) = \prod_{j=1}^{N_{\rm bins}} \frac{\mu_j^{d_j} e^{-\mu_j }}{ d_j!}, 
\label{eq:PoissonLikelihood}
\end{equation}
where in the special case of a constant-rate hypothesis we take $\mu_j  = \mu$ for all $j$, 
i.e., ${\bm \mu} = \{\mu, \mu, \ldots, \mu\}$ and
\begin{align}
\mathcal L(\vec d | \bm{\mu}) &= \mu^N e^{- N_{\rm bins}\mu} \prod_{j=1}^{N_{\rm bins}} \frac{1}{d_j!} . 
\label{eq:poisconst}
\end{align}

We define the $p$-value $p(\vec d | \vecb\mu)$ as
\begin{align}
p(\vec d | \vecb\mu) &\equiv \sum_{\vec d'} \mathcal L(\vec d' | \vecb{\mu}) \times \Theta\big( \mathcal L(\vec d' | \vecb{\mu})  \leq  \mathcal L(\vec d | \vecb{\mu}) \big), 
\label{eq:pval}
\end{align}
where $\Theta( x \leq y)$ is the Heaviside $\Theta$ function $\Theta(y - x)$ defined with $\Theta(0) \equiv 1$. A small value for $p(\vec d | \vecb\mu)$ indicates that it is unlikely that the data $\vec d$ were generated from the Poisson distribution parameterized by $\vecb\mu$.
For convenience, a $p$-value can be converted into a ``number of sigmas'' $N_\sigma$ via
\begin{align}
N_\sigma(p) &\equiv \sqrt{2} \erf^{-1}\left( 1 - p \right) ,
\label{eq:nsigma}
\end{align}
so that $1 - p = \{68.27\%, 95.45\%, 99.73\%, \ldots\}$ map respectively onto $\{1\sigma, 2\sigma, 3\sigma, \ldots\}$. 
The best-fit hypothesis is the one that maximizes $\mathcal{L}(\vec d | \vecb\mu)$. 

It is also convenient to define a test statistic $q$ as a monotonic function of $\mathcal L$, 
\begin{align}
q(\vec d | \vecb\mu) &\equiv -2 \log \mathcal L(\vec d | \vecb\mu).
\label{eq:TS}
\end{align}
Because $q(\mathcal L)$ decreases monotonically, the $p$-value can be equally well expressed as 
\begin{align}
p(\vec d | \vecb\mu) &\equiv \sum_{\vec d'} \mathcal L(\vec d' | \vecb{\mu}) \times \Theta\big( q(\vec d' | \vecb{\mu})  \geq  q(\vec d | \vecb{\mu}) \big).
\label{eq:qval}
\end{align}
Using the Stirling approximation for $\log x!$, the exact form of $q(\vec d | \vecb \mu)$ is well approximated by
\begin{align}
q(\vec d | \vecb \mu) &\simeq \sum_j \left \{ 2(\mu_j - d_j) + 2d_j \log \frac{d_j}{\mu_j} + \log(2 \pi d_j) \right \},
\label{eq:qStirling}
\end{align}
as long as $\mu_j \gtrsim 1$ and $d_j \geq 1$ for all $j$. 

\subsubsection{Gaussian Limit}
In the limit of large $\mu_j \gg 10$, we can derive a much simpler expression for $p(\vec d | \vecb\mu)$ by approximating $\mathcal L$ as a Gaussian probability distribution:
\begin{align}
\mathcal L(\vec d | \vecb\mu) &\approx \prod_j \frac{1}{\sqrt{ 2 \pi \mu_j} } \exp\left( - \frac{(d_j - \mu_j)^2 }{2 \mu_j } \right),
\\
q(\vec d | \vecb\mu)  = - 2 \log \mathcal L & \approx \sum_j \left \{\log( 2 \pi \mu_j) + \frac{(d_j - \mu_j)^2 }{\mu_j} \right\}  . 
 \label{eq:qGauss}
\end{align}
Note that \eqref{eq:qGauss} can also be derived from \eqref{eq:qStirling}, by taking a series expansion about $\vec d \approx \bm{\mu} + \mathcal O(\sqrt{\bm{\mu}})$. The $\vec d$-dependent term in \eqref{eq:qGauss} is also known as the $\chi^2$ distance between vectors $\vec d$ and $\vecb \mu$:
\begin{align}
\chi^2(\vec d | \vecb\mu) &\equiv  \sum_{j} \frac{(d_j - \mu_j)^2 }{\mu_j} .
\end{align}
The corrections to the Gaussian version of $q(\vec d | \vecb\mu)$ scale as $1/\sqrt{\bm{\mu}}$, making \eqref{eq:qGauss} sufficiently accurate in the $\bm{\mu} \gg 10$ limit, provided that each $d_j$ is within $d_j \approx \mu_j \pm 3 \sqrt{\mu_j}$ of the mean.

When evaluating the $p$-value, $p(\vec d | \vecb\mu)$, the $\log \bm{\mu}_j$ terms in \eqref{eq:qGauss} can be ignored: 
\begin{align}
p(\vec d | \vecb\mu) &\approx \sum_{\vec d'} \mathcal L(\vec d' | \vecb{\mu}) \times \Theta\big( \chi^2(\vec d' | \vecb{\mu})  \geq  \chi^2(\vec d | \vecb{\mu}) \big).
\end{align}
In the limit of large $\bm{\mu}$, the discrete sum can also be approximated by an integral:
\begin{align}
p(\vec d | \vecb\mu) &\approx \int \!d{\bm{d}}' \mathcal L(\vec d' | \vecb\mu) \times \Theta\big(\chi^2(\vec d' | \vecb \mu ) - \chi^2(\vec d | \vecb \mu)\big)
\\
&= \int\! d {\bf u} \, \Theta\big(|\vec u|^2 - \chi^2(\vec d | \vecb\mu) \big) \prod_j \frac{\exp(- u_j^2 /2 ) }{\sqrt{ 2\pi} }  ,
\end{align}
which can be converted to spherical coordinates such that
\begin{align}
1 - p(\vec d | \vecb\mu) &\approx \frac{1}{\Gamma\left( \frac{N_{\rm bins}}{2} \right) } \int_0^{\sqrt{ \chi^2(\vec d|\vecb\mu) } } \! u^{N_{\rm bins}-1} du\, e^{-u^2/2} 
\nonumber\\&= \frac{\gamma\left( \frac{N_{\rm bins}}{2}, \frac{1}{2} \chi^2(\vec d | \vecb\mu) \right) }{\Gamma\left( \frac{N_{\rm bins}}{2} \right) } , 
\end{align}
where $\gamma(a, z)$ is the lower incomplete gamma function. 
Equivalently, from \eqref{eq:nsigma}, 
\begin{align}
N_\sigma(\vec d | \vecb\mu) &\approx \sqrt{2} \erf^{-1}\left(  \frac{\gamma\!\left( \frac{N_{\rm bins}}{2}, \frac{1}{2} \chi^2(\vec d | \vecb\mu) \right) }{\Gamma\left( \frac{N_{\rm bins}}{2} \right) } \right) .
\label{eq:nsigmagaussian}
\end{align}
Both $p$ and $N_\sigma$ are monotonic in $\chi^2(\vec d | \vecb\mu)$. 

\subsubsection{Flat background} \label{sec:flat}
Let us specialize to the case where the background is constant in time. Here, the background model is parameterized by a single constant: $R_\text{bkg}$, the average background rate. The average number of events expected in the $j$th bin is simply:
\begin{align}
\mu_j &= t_\text{bin} \times R_\text{bkg}, 
\end{align}
which is identical for all bins if the exposure times $t_\text{bin}$ are equal. 
It is trivial to show from \eqref{eq:poisconst} that $\mathcal L (\vec d | \bm{\mu})$ is maximized at a $\bm{\mu} = \bm{\mu}_0$ satisfying 
\begin{equation}
\mu_j = \mu_0 = \frac{N}{N_{\rm bins}},\qquad R_\text{bkg} = \frac{\mu_0}{t_\text{bin} } =  \frac{N }{T}, 
\end{equation}
where $T = N_{\rm bins} t_\text{bin}$ is the total exposure time. 
We describe this $R_\text{bkg}$  as the ``measured average rate.''

Maximizing $\mathcal L(\vec d | \bm{\mu})$ is not exactly the same as minimizing $\chi^2(\vec d | \bm{\mu})$ with respect to $\bm{\mu}$. 
The value of $\mu$ that does the latter is actually
\begin{align}
\mu_\text{fit} &= \sqrt{ \frac{1}{N_{\rm bins}} \sum_{j = 1}^{N_{\rm bins}} d_j^2 }. 
\end{align}
The difference between $\mu_\text{fit}$ and $\mu_{0}$ depends on the size of the fluctuations in the dataset. 
In the Gaussian limit where \eqref{eq:qGauss} is appropriate, however, we do expect $\chi^2(\vec d| {\bm \mu}_0) \approx\chi^2(\vec d| {\bm \mu}_\text{fit})$.
To take an example with $\mu \sim 100$, the two versions of ``best fit'' background rates give values for $N_\sigma$ that differ by less than 0.1 even for $5 \sigma$ fluctuations from the mean. Generally, if the distinction between $\mu_0$ and $\mu_\text{fit}$ becomes important to the analysis, then the Gaussian limit was not sufficiently accurate.

Finally, the total exposure $N$ can be factored out of the expression for $\chi^2$ as follows: 
\begin{align}
\chi^2(\vec d | \bm{\mu}_0) &= (N_{\rm bins} \mu_0) \frac{1}{N_{\rm bins} } \sum_j \left( \frac{d_j }{\mu_0} - 1 \right)^2 \\
&= N \times \frms^2(\vec d | \bm{\mu}_0), 
\end{align}
where (still assuming a flat background)
\begin{align}
\frms(\vec d | \bm{\mu}) &\equiv  \sqrt{\frac{1}{N_{\rm bins}} \sum_{j} \left( \frac{d_j}{\mu} - 1 \right)^2 }. 
\label{def:frmsdiscrete}
\end{align}
Here $\frms$ is the fractional root-mean-square amplitude of the measured signal $\vec d$, relative to the constant $\bm{\mu}$. 
To maximize the statistical strength of a fluctuating signal, then, the quantity we want to maximize is 
\begin{align}
N_\sigma(\vec d |\bm{\mu}) \sim \sqrt{\chi^2(\vec d | \bm{\mu})} &\simeq  \sqrt{N}  \times  \frms(\vec d | \bm{\mu}) .
\label{eq:main}
\end{align}
This is one of our primary results.

A value of $N_\sigma \lesssim 2$ indicates that the data are well described by a constant background rate across bins, $R_\text{bkg} = \mu/t_\text{bin}$. 
A larger value of $N_\sigma \gtrsim 3$ is not specifically evidence for a dark matter signal, however. For example, non-Poissonian noise with a standard deviation much larger than $\sqrt{{\bm \mu}}$ would also produce a large $N_\sigma$.

The same RMS modulation amplitude also governs the discovery and exclusion test statistics. The null model of no DM, corresponding to a fixed $A = 0$, is nested within the signal model in which $A$ is allowed to take nonzero values. We can then define a frequentist test statistic for these nested hypotheses as
\begin{equation}
    q(\vec d| A) = 2\log \left[\frac{\mathrm{max}_{\bm{\Theta}}\mathcal{L}(\bm{d}|A, \bm{\Theta})}{\mathrm{max}_{\bm{\Theta}}\mathcal{L}(\bm{d}|A=0, \bm{\Theta})} \right].
    \label{eq:TSqA}
\end{equation}
Evaluating the test statistic $q(A)$ and its expected distribution under the null hypothesis (e.g., $A = 0$) can be time-consuming. This evaluation may require either detailed analytics (possible only in special cases) or a large amount of well-calibrated Monte Carlo simulation. Nevertheless, analytic progress can be made in the asymptotic limit of a large number of counts~\cite{Wilks:1938dza, 60455556-c6b2-37bb-9077-b6ce7e1443d8}.

In particular, define the expected modulation fraction $f_j$ in each bin and the observed modulation fraction $\delta_j$ as
\begin{equation}
1 + f_j \equiv \frac{A r_j + b_s(A)}{\mu_0} ,\qquad
1 + \delta_j \equiv \frac{d_j}{ \mu_0} .
\end{equation}
Here $b_s(A)$ is the best-fit background rate for the signal model parameterized by $A$. It is one of the parameters in $\vecb\Theta$ over which $\mathcal L(\vec d | A, \vecb\Theta)$ is profiled.
In the Gaussian limit, the test statistic is
\begin{align}
q(\vec d| A)  &\simeq \sum_j \left \{\mu_0 (\delta_j)^2 - \frac{\mu_0}{1 + f_j} \left( \delta_j - f_j \right)^2  - \log(1 + f_j) \right \} .
\label{eq:qFracs}
\end{align}
First, consider data that match the flat-background distribution, $d_j \simeq \mu_0$, in which case $\delta_j \simeq 0$. In the weak-signal limit, we can expand to leading order in $f_j \ll 1$. Dropping the log term, which does not scale with $\mu_0 \propto N$, we have
\begin{align}
q(\vec d \simeq {\bm \mu}_0 | A) &\approx - \mu_0 \sum_j f_j^2   
\approx - N \times f_\text{RMS}^2( \vecb\mu(A)  | {\bm \mu}_0), 
\end{align}
where we have used the definition of $f_{\rm RMS}$ from \eqref{def:frmsdiscrete}. 
Next, consider a data set that matches the DM model quite well, i.e., $\delta_j \simeq f_j$. 
Dropping the small $(\delta_j - f_j)^2$ term and the non-extensive $\log(1 + f_j)$ term from \eqref{eq:qFracs}, we find 
\begin{align}
q(\vec d \simeq \vecb\mu_s | A) &\simeq \mu_0 \sum_j (\delta_j)^2 = N \frac{1}{N_{\rm bins}} \sum_j \left( \frac{d_j}{\mu_0} - 1 \right)^2 
\\
&\simeq N \times f_\text{RMS}^2(\vec d | {\bm \mu}_0 ) ,
\end{align}
which also scales like \eqref{def:frmsdiscrete}. We thus see that in the Gaussian limit of a large number of counts per bin and a flat background, both the exclusion and discovery significance will generically scale like $\sqrt{N} \times f_{\rm RMS}$, identifying $f_{\rm RMS}$ as the appropriate figure of merit for daily modulation significance.

\subsection{Fisher information derivation} \label{sec:fisherinfo}
Here we present an alternative derivation of the RMS modulation amplitude as the figure of merit, using the Fisher information. As in the previous section, we take the bins to be of uniform size $t_{i+1} - t_i = t_{\rm bin}$, with $T = N_\mathrm{bins} t_{\rm bin}$. The $d_i$ then each follow a Poisson distribution with mean
\begin{equation}
    \mu_i(A, \bm{\Theta}) = \int_{t_i}^{t_{i+1}} dt\, r(t| \{A, \bm{\Theta}\}) \equiv A S_i + B_i(\bm{\Theta})
\end{equation}
where we have defined
\begin{equation}
S_i = \int_{t_i}^{t_{i+1}} dt \, s(t),\quad 
B_i(\bm{\Theta}) = \int_{t_i}^{t_{i+1}} dt \, b(t|\bm{\Theta}).
\end{equation}
Since the data are drawn from a Poisson process, the Poisson likelihood in Eq.~(\ref{eq:PoissonLikelihood}) is the appropriate one for inference or parameter estimation. We see here the utility of our assumption of a small signal: our Poisson means never become negative, and thus the Poisson likelihood remains well-defined.

In the asymptotic limit, the variance of the maximum likelihood estimate of the signal-strength parameter can be used to determine projected limits under the null or expected detection thresholds~\cite{Cowan:2010js}. We assume that the ``true'' model describing a hypothetical dataset $\vec d$ is specified by $A^t$ and $\bm{\Theta}^t$, so that the Fisher information matrix for asymptotically unbiased estimators is given by
\begin{equation}
    \mathcal{I} = \begin{bmatrix}
        \mathcal{I}_{AA} & \mathcal{I}_{\bm{\Theta} A} \\
        \mathcal{I}_{A \bm{\Theta}} & \mathcal{I}_{\bm\Theta\bm\Theta}
    \end{bmatrix}
\end{equation}
where the matrix elements are defined as the scalar
\begin{equation}
    \mathcal{I}_{AA} = -\mathbb{E}\left[\frac{\partial}{\partial A} \log \mathcal{L}(\mathbf{d} |A, \bm{\Theta})\, \frac{\partial}{\partial A}\log\mathcal{L}(\mathbf{d} |A, \bm{\Theta}) \right]_{A^t, \bm{\Theta}^t}
\end{equation}
the vector
\begin{equation}
    [\mathcal{I}_{A \bm{\Theta}}]_{i} =  -\mathbb{E}\left[\frac{\partial}{\partial A} \log \mathcal{L}(\mathbf{d} |A, \bm{\Theta}) \, \frac{\partial}{\partial \Theta_i}\log\mathcal{L}(\mathbf{d} |A, \bm{\Theta}) \right]_{A^t, \bm{\Theta}^t},
\end{equation}
and the matrix 
\begin{equation}
    [\mathcal{I}_{ \bm{\Theta} \bm{\Theta}}]_{ij} =  -\mathbb{E}\left[\frac{\partial}{\partial \Theta_i} \log \mathcal{L}(\mathbf{d} |A, \bm{\Theta}) \, \frac{\partial}{\partial \Theta_j}\log\mathcal{L}(\mathbf{d} |A, \bm{\Theta}) \right]_{A^t, \bm{\theta}^t}.
\end{equation}
For our Poissonian likelihood, working at leading order in the small-signal limit, these quantities simplify to 
\begin{equation}
\begin{gathered}
\mathcal{I}_{AA} = \sum_{k=1}^{N_{\rm bins}} \frac{S_k^2}{B_k}, \qquad 
[\mathcal{I}_{A\bm{\Theta}}]_i = \sum_{k=1}^{N_{\rm bins}} \frac{S_k}{B_k} \frac{\partial B_k}{\partial \Theta_i} \\
[\mathcal{I}_{\bm\Theta\bm\Theta}]_{ij} =\sum_{k=1}^{N_{\rm bins}} \frac{1}{B_k}\frac{\partial B_k}{\partial \Theta_i}\frac{\partial B_k}{\partial \Theta_j}
\end{gathered}
\end{equation}
where $\vec B$ is evaluated at $\bm{\Theta}^t$. From the Fisher information matrix, we can calculate the variance of the maximum-likelihood estimator for $A$ via the Schur complement, for the data $\vec d$ generated under the true model by
\begin{equation}
    \sigma_A^{-2}  =  \mathcal{I}_{AA} - \mathcal{I}_{A \bm{\Theta}} \mathcal{I}^{-1}_{\bm{\Theta}{\bm{\Theta}}} \mathcal{I}_{\bm{\Theta}A}.
    \label{eq:FisherVariance}
\end{equation}
in terms of the inverse of the $\mathcal{I}_{\bm{\Theta} \bm{\Theta}}$ submatrix. Hence we can see that, at fixed $\mathcal{I}_{AA}$, the most precise (i.e.\ smallest-variance) estimator is obtained when $\mathcal{I}_{A \bm{\Theta}} = 0$ which occurs when the signal template $\vec S$ is orthogonal to any gradient of the background template $\vec B$ with respect to the nuisance parameters. Physically, this can be interpreted as a case where there is no degeneracy between the signal and background models.

If the background model is specified by multiple parameters, then $\mathcal{I}$ is high-dimensional and evaluating the estimator variance from Eq.~(\ref{eq:FisherVariance}) becomes analytically intractable but remains straightforward to perform numerically. However, an informative and simple case is one in which the background rate is specified by a single parameter, $b(t) \propto \Theta_B$. A particularly convenient parametrization is
\begin{equation}
    b(t) = B_\mathrm{tot} \hat{b}(t) \quad\mathrm{for}\quad \int_0^T dt \,\hat{b}(t) = 1
\end{equation}
where the unknown background parameter is now the total number of expected background events over the measurement period, $\Theta_B \equiv B_\mathrm{tot}$. This analysis is appropriate for background models where the profile $\hat b(t)$ is known, but its amplitude (the rate of background events) is not. For this convenient choice, we have
\begin{equation}
    \frac{\partial \vec B}{\partial B_\mathrm{tot}} = \frac{1}{B_\mathrm{tot}} \vec B,\qquad \hat{\vec B} = \frac{\vec B}{B_\mathrm{tot}}.
\end{equation}
Similarly, we can define 
\begin{equation}
\begin{gathered}
    S_\mathrm{tot} = \int_0^T dt\, s(t), \qquad \hat{s}(t) = \frac{s(t)}{S_\mathrm{tot}} \\
    \hat{\vec S} = \frac{\vec S}{S_\mathrm{tot}}
\end{gathered}
\end{equation}
so that we obtain
\begin{equation}
\begin{split}
    \sigma_A^{-2} &= \frac{S_\mathrm{tot}^2}{B_\mathrm{tot}} \left[ \sum_k  \frac{\hat{S}_k^2}{\hat{B}_k} -1\right]
\end{split}
\end{equation}
We can also take this to the continuum limit (which we further justify in Sec.~\ref{sec:Binning} below), where we obtain 
\begin{equation}
\begin{split}
    \sigma_A^{-2} \approx \frac{S_\mathrm{tot}^2}{B_\mathrm{tot}} \left[ \left( \int_0^T dt  \frac{\hat{s}(t)^2}{\hat{b}(t)}\right) -1\right].
\end{split}
\end{equation}
These asymptotic results enable us to identify the detection figure-of-merit, representing the root-mean-square variation in signal as a function of time, as 
\begin{equation}
f_\mathrm{RMS}^2 \equiv
\begin{cases}
    \sum_k  \dfrac{\hat{S}_k^2}{\hat{B}_k} -1, & \text{(discrete)}, \\[2.5ex]
    \int dt  \dfrac{\hat{s}(t)^2}{\hat{b}(t)} -1 , & \text{(continuum) }.
\end{cases} 
\label{eq:fRMS_chi2}
\end{equation}
In the even simpler case where the background rate is constant, we have $\hat{b}(t) = 1/T$. Defining the normalized rates
\begin{equation}
    r(t) \equiv \frac{s(t)}{\int_0^T \frac{dt}{T} s(t)}, \qquad r_i = \int_{t_i}^{t_{i+1}} \!dt\,  r(t),
\end{equation}
we obtain
\begin{equation}
f_\mathrm{RMS}^2 \equiv
\begin{cases}
    \frac{1}{N_\mathrm{bins}} \sum_k r_k^2 - 1 & \text{(discrete)}, \\[1.5ex]
     \int  \frac{dt}{T} r(t)^2  -1 , & \text{(continuum) }
\end{cases} 
\end{equation}
as anticipated in Eq.~(\ref{eq:frms}) for the continuum case and Eq.~(\ref{def:frmsdiscrete}) for the discrete case. 
Moreover, up to intra-day modulation, $S_\mathrm{tot}$ and $B_\mathrm{tot}$ scale linearly with the observation time $T$, and so we have 
\begin{equation}
    \sigma_A^{-2} = \frac{S_\mathrm{tot}^2}{B_\mathrm{tot}}f_\mathrm{RMS}^2 \sim T f_\mathrm{RMS}^2.
    \label{eq:significance}
\end{equation}
The discovery significance, proportional to $1/\sigma_A$, thus scales as $f_\mathrm{RMS} \sqrt{T}$. In the sections which follow, we show that this result generalizes to configurations of multiple detectors (Sec.~\ref{sec:MultiLikelihood}) and arbitrary time binning (Sec.~\ref{sec:Binning}).
\section{Multiple Detector Likelihood}
\label{sec:MultiLikelihood}
To consider an ensemble of $N_\mathrm{det.}$ detectors, we can straightforwardly extend Eq.~(\ref{eq:PoissonLikelihood}) to 
\begin{equation}
    \mathcal{L}(\bm{d} | \{ A,  \bm{\Theta}\})=     \prod_{i=1}^{N_\mathrm{det.}}\prod_{j=1}^{N_\mathrm{bins}}\frac{\mu_{i,j}^{d_{i,j}} e^{-\mu_{i,j}}}{d_{i,j}!},
    \label{eq:MultiplePoissonLikelihood}
\end{equation}
where $d_{i,j}$ is the observed number of events in the $j^\mathrm{th}$ bin of the $i^\mathrm{th}$ detector. Likewise,  $\mu_{i,j}$ is the expected number of events in the $j^\mathrm{th}$ bin of the $i^\mathrm{th}$ detector, with 
\begin{equation}
    \mu_{i,j} = A S_{i,j} + B_{i,j}(\bm{\Theta})
    \label{eq:mu_mult_det}
\end{equation}
where we have defined
\begin{equation}
S_{i,j} = \int_{t_i}^{t_{i+1}} dt \, s_i(t),\quad 
B_{i,j}(\bm{\Theta}) = \int_{t_i}^{t_{i+1}} dt \, b_i(t|\bm{\Theta}).
\label{eq:SB_mult_det}
\end{equation}
for signal rate $s_i(t)$ and background rate $b_i(t)$ at the $i^\mathrm{th}$ detector. 

We can most straightforwardly make contact with our previous asymptotic results by block-embedding all the relevant vectors so that we have
\begin{equation}
\begin{gathered}
    \bar{\vec d} = \begin{bmatrix}
        {\vec d}_1^T & {\vec d}_2^T & \hdots & {\vec d}_{N_\mathrm{det.}}^T 
    \end{bmatrix}^T \\
    \bar{\bm{\mu}} = \begin{bmatrix}
        \bm{\mu}_1^T & \bm{\mu}_2^T & \hdots & \bm{\mu}_{N_\mathrm{det.}}^T 
    \end{bmatrix}^T \\
    \bar{\vec S} = \begin{bmatrix}
        {\vec S}_1^T & {\vec S}_2^T & \hdots & {\vec S}_{N_\mathrm{det.}}^T 
    \end{bmatrix}^T \\
    \bar{\vec B} = \begin{bmatrix}
        {\vec B}_1^T & {\vec B}_2^T & \hdots & {\vec B}_{N_\mathrm{det.}}^T 
    \end{bmatrix}^T .
\end{gathered}
\end{equation}
The likelihood now can be written
\begin{equation}
    \mathcal{L}(\bar{\bm{d}} | \{ A,  \bm{\Theta}\})= \prod_{a=1}^{N_\mathrm{bins} \times N_{\rm det.}}\frac{\bar{\mu}_a^{\bar{d}_a} e^{-\bar{\mu}_a}}{\bar{d}_a!}.
    \label{eq:MultiPoissonLikelihood}
\end{equation}
with information matrix
\begin{equation}
    \mathcal{I} = \begin{bmatrix}
        \mathcal{I}_{AA} & \mathcal{I}_{A \bm{\Theta}}^T \\
        \mathcal{I}_{A \bm{\Theta}} & \mathcal{I}_{\bm\Theta\bm\Theta}
    \end{bmatrix}
\end{equation}
with 
\begin{equation}
\begin{gathered}
\mathcal{I}_{AA} = \sum_{a=1}^{N_{\rm bins}\times N_{\rm det.}}  \frac{\bar{S}_a^2}{\bar{B}_a}, \qquad 
[\mathcal{I}_{A\bm{\Theta}}]_i = \sum_{a=1}^{N_{\rm bins}\times N_{\rm det.}} \frac{\bar{S}_a}{\bar{B}_a} \frac{\partial \bar{B}_a}{\partial \Theta_i} \\
[\mathcal{I}_{\bm\Theta\bm\Theta}]_{ij} =\sum_{a=1}^{N_{\rm bins}\times N_{\rm det.}} \frac{1}{\bar{B}_a}\frac{\partial \bar{B}_a}{\partial \Theta_i}\frac{\partial \bar{B}_a}{\partial \Theta_j},
\end{gathered}
\label{eq:Fisher_mult_det}
\end{equation}
a clear generalization of our previous single-detector result. We then obtain $\sigma_A^{-2}$ from $\mathcal{I}$ using the construction in Eq.~(\ref{eq:FisherVariance}).

\subsection{Sensitivities for independent backgrounds}
\label{sec:UncorrelatedBkg}
The information matrix simplifies even further if the background rate at each detector is independently parametrized, as might be expected for an endogenous background rate for each detector material (for example, proportional to the carbon-14 content) in an array of such detectors. This gives $\mathcal{I}_{\bm{\Theta}\bm{\Theta}}$ a block-diagonal structure. In particular, if the background at the  $i^\mathrm{th}$ detector is specified by $\bm{\Theta}^{(i)}$, then we ultimately find
\begin{equation}
\sigma_A^{-2} =  \sum_{i=1}^{N_{\rm det.}} \left[\mathcal{I}_{AA}^{(i)}- \mathcal{I}_{A\bm{\Theta}}^{(i)} \left( \mathcal{I}_{\bm{\Theta}\bm{\Theta}}^{(i)} \right)^{-1} \mathcal{I}_{\bm{\Theta} A}^{(i)} \right]
\end{equation}
where we have defined the detector-indexed components of the information matrix by
\begin{equation}
\begin{gathered}
\mathcal{I}_{AA}^{(i)} = \sum_{k=1}^{N_{\rm bins}}  \frac{S_{i, k}^2}{B_{i, k}}, \qquad 
[\mathcal{I}_{A\bm{\Theta}}^{(i)}]_j = \sum_{k=1}^{N_{\rm bins}} \frac{S_{i, k}}{B_{i,k}} \frac{\partial B_{i,k}}{\partial \Theta^{(i)}_j} \\ \\
[\mathcal{I}_{\bm\Theta\bm\Theta}^{(i)}]_{jl} =\sum_{k=1}^{N_{\rm bins}}  \frac{1}{B_{i, k}}\frac{\partial B_{i, k}}{\partial \Theta^{(i)}_j}\frac{\partial B_{i, k}}{\partial \Theta^{(i)}_l}.
\end{gathered}
\end{equation} 
Keeping careful track of the indices, we can recognize that the information matrix for the $i^\mathrm{th}$ detector is 
\begin{equation}
    \mathcal{I}^{(i)} = \begin{bmatrix}
        \mathcal{I}_{AA}^{(i)} & (\mathcal{I}_{A\bm{\Theta}}^{(i)})^T \\
        \mathcal{I}_{A\bm{\Theta}}^{(i)} & \mathcal{I}_{\bm{\Theta}\bm{\Theta}}^{(i)}
    \end{bmatrix},
\end{equation}
so that 
\begin{equation}
    \sigma_A^{-2} = \sum_{i=1}^{N_{\rm det.}} \left[\sigma_A^{(i)}\right]^{-2}
\end{equation}
where $\left[\sigma_{A}^{(i)}\right]^2$ is the variance in the signal strength estimator. This recovers the expected result that sensitivities add harmonically in quadrature and that the best sensitivity is achieved when each detector is operated in a manner that maximizes its independent sensitivity, providing no opportunities for further optimization.

\subsection{Sensitivities for correlated backgrounds}
\label{sec:correlated}

More nontrivial optimal detection strategies are possible when backgrounds are correlated across multiple detectors, as might be expected from an extrinsic background source. While we consider some more complicated toy models in greater detail in Sec.~\ref{sec:Example}, we present a simplified scenario here to build intuition. Specifically, we consider two detectors which have an identical background rate $b(t | \theta)$ specified by a single parameter $\theta$. In the case where we have identical signals ${\vec S}_1 = {\vec S}_2 = {\vec S}$, then we obtain
\begin{equation}
\begin{split}
    \frac{\sigma_A^{-2}}{2} &=   \sum_{j=1}^{N_{\rm bins}} \frac{S_{j}^2}{B_j} - \left( \sum_{j=1}^{N_{\rm bins}}\frac{S_{j} \partial_\theta B_j}{B_j} \right)^{\! 2} \!\! \left(\sum_{k=1}^{N_{\rm bins}} \frac{(\partial_\theta B_{k})^2}{B_k} \right)^{\!-1}.
\end{split}
\label{eq:multicorrel}
\end{equation} 
At the other extreme, consider the case ${\vec S}_1 = -{\vec S}_2 = {\vec S}$, which could be realized for an oscillatory signal which is completely out of phase at detector 2 as compared to at detector 1.\footnote{Recall that we are always in the weak signal limit, so the signal counts never actually go negative; moreover, any constant component of the signal is fully degenerate with a constant background event rate so can be reparametrized in a matter such that it is absorbed into the inferred background rate.} In that case, the sensitivity is
\begin{equation}
\begin{split}
    \sigma_A^{-2} &= 2 \sum_{j=1}^{N_{\rm bins}} \frac{S_{j}^2}{B_j},
\end{split}
\end{equation} 
which is strictly greater than the sensitivity realized for ${\vec S}_1 = {\vec S}_2$ when there is degeneracy between the signal and background measured at a single detector.

The reason for the improvement is that ${\vec S}_1 = -{\vec S}_2$ reduces the overlap between the total signal vector and the total background gradient vector while leaving the signal component of the information, $\mathcal{I}_{AA}$ (which depends only on $S_i^2$), unchanged. While physically relevant scenarios may not so easily allow for optimal signal alignment like ${\vec S}_1 = -{\vec S}_2$, this example illustrates that detector parameters which modulate the relative amplitude and phase of a signal at a given detector can result in improved sensitivity if chosen carefully, even though they may have no impact on the estimated single-detector sensitivity.

\section{Impact of Time Binning and Validity of Asymptotics}
\label{sec:Binning}

In this section, we validate our application of asymptotic results to situations where varying the time binning may result in smaller or larger populations of events in each bin. Perhaps surprisingly, it is \emph{not} true that the asymptotic results will fail even in binned analyses when the expected number of events in a given bin becomes small and high-variance due to Poisson fluctuations. To the contrary, asymptotic results can remain valid even if the expected number of events per bin becomes vanishingly small, \emph{so long as the log-likelihood and score function remain approximately Gaussian}. Indeed, the log-likelihood will generally take the form of a sum over data, either binned or unbinned, and so long as many independent summands make approximately equal contributions to the total log-likelihood, the central limit theorem implies its approximate Gaussianity.

\begin{figure*}[!ht]
\centering
\includegraphics[width=0.99\textwidth]{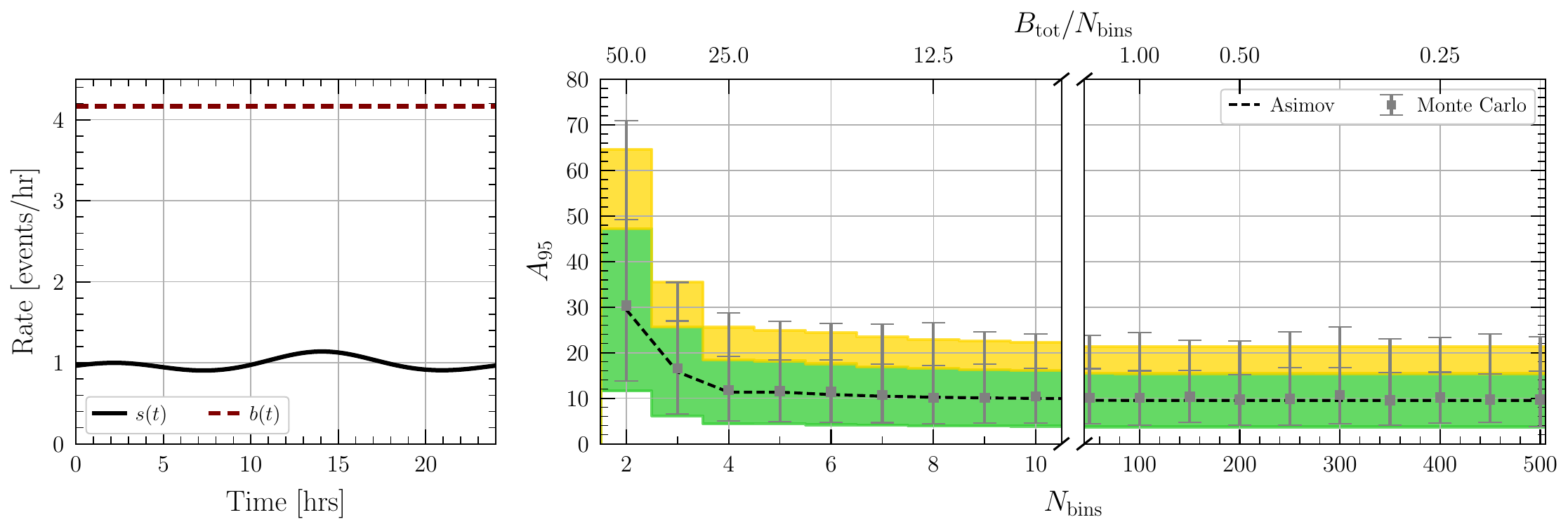}
\caption{(\textit{Left}) A comparison of signal and background templates used in our Monte Carlo data-generation procedure. In black, we depict the signal rate $s(t)$ for $A = 1$ while in dashed red, we depict the uniform background rate for $B_\mathrm{tot} = 100$. (\textit{Right}) A comparison of the asymptotic sensitivity estimates with the results of the Monte Carlo procedure for generating limits under the null as a function of the number of bins used to resolve the total event rate jointly produced by signal and background processes. As a function of $N_\mathrm{bins}$, the black dashed line indicates the median expected $A_{95}$ in the asymptotic limit, while the interval in green represents the central 68\% confidence interval for $A_{95}$ under the null; the interval in yellow represents the $84^\mathrm{th}$ to $97.5^\mathrm{th}$ percentile values of $A_{95}$ in its expected distribution under the null. The grey data points indicate the median of a Monte Carlo sampling of $A_{95}$ for data generated under the null, while the lower error bar (first upper error bar) [second upper error bar] indicates the 16$^\mathrm{th}$ (84$^\mathrm{th}$) [97.5$^\mathrm{th}$] percentile value in the sampled distribution. See text for the details of this construction. The relatively good agreement across $N_\mathrm{bins}$ indicates that our asymptotic sensitivity estimates already are trustworthy in these instances with only $\sim 100$ events in the dataset, with correspondingly better convergence expected for datasets with more events. We emphasize that the level of agreement is constant across $N_\mathrm{bins}$, indicating that our results hold in the narrow or even unbinned limit.}
\label{fig:monte_carlo_demonstration}
\end{figure*}

Assessing whether or not asymptotic results can be reliably applied to a given dataset or projection must, of course, be performed on a case-by-case basis and will depend on the desired statistical power. The purpose of this section is to demonstrate empirically that the results derived in this work can remain valid even in the very small bin-occupancy limit, which we validate here through the study of a representative set of Monte Carlo data analyses. In practice, this means that a daily modulation analysis is free to choose any convenient binning scheme, which may depend on the experimental demands of the apparatus. 

\subsection{Asymptotic Projections}

In the left panel of Fig.~\ref{fig:monte_carlo_demonstration}, we illustrate our model components, which we consider in the context of data collection with a single detector over the course of a single day. We adopt a representative signal profile from a trans-stilbene detector~\cite{Blanco:2021hlm} oriented such that the signal rate realizes a roughly 25\% peak-to-peak variation over a 24-hour period.\footnote{For the purposes of this illustration, the distinction between Earth day and sidereal day is not relevant; we properly implement the sidereal modulation in Secs.~\ref{sec:Sidereal} and \ref{sec:Example}.} We also include a uniform background rate. The data are thus generated according to the model\footnote{In this section, we do not attempt to relate the signal model to realistic experimental parameters such as exposure, DM mass, and DM cross section; we leave such an analysis for Sec.~\ref{sec:Example}.}
\begin{equation}
    r(t) = A\,s(t) + \frac{B_\mathrm{tot}}{24\,\mathrm{hr}}.
\end{equation}
To assess our limit-setting sensitivity as a function of our binning, we choose ``true" underlying model values $A^t = 0$ and $B_\mathrm{tot}^t = 100$. Then, using the machinery developed in Sec.~\ref{sec:SingleLikelihood}, we determine the asymptotically expected $\sigma_A$, as a function of the number of time-bins, $N_\mathrm{bins}$, that uniformly resolve the 24 hour data-taking period. In the asymptotic limit, the median frequentist $95^\mathrm{th}$ percentile upper limit on the signal strength parameter is given by 
\begin{equation}
    A_\mathrm{95} = \Phi^{-1}(0.95) \times \sigma_A
\end{equation}
where $\Phi^{-1}$ is the percentile point function of the standard normal distribution. Similarly, the $1\sigma$ confidence interval for the 95$^\mathrm{th}$ percentile limit is 
\begin{equation}
    A_\mathrm{95}^\pm = [\Phi^{-1}(0.95) \pm1] \times \sigma_A,
\end{equation}
while the $2\sigma$ upper limit for the 95$^\mathrm{th}$ percentile limit is
\begin{equation}
     A_\mathrm{95}^{++} = [\Phi^{-1}(0.95) + 2] \times \sigma_A.
\end{equation}
For more details, see, \textit{e.g.}, \cite{Cowan:2010js}. We do not calculate a $2\sigma$ lower limit for the 95$^\mathrm{th}$ percentile upper limit, since typically limits are power-constrained at the $1\sigma$ lower limit~\cite{Cowan:2011an}. We perform these asymptotic sensitivity estimates for a representative set of values for $N_\mathrm{bins}$ ranging between $2$ and $500$. Of course, for $N_\mathrm{bins} = 1$, the signal and background become fully degenerate.

\subsection{Monte Carlo Sensitivity Expectations}
To evaluate the accuracy of our asymptotic sensitivity estimates, we perform Monte Carlo simulations that enable direct construction of the median 95$^\mathrm{th}$ upper limit and its associated containment interval. In our Monte Carlo simulation procedure, for a given $N_\mathrm{bins}$, we generate a null-model data realization $A^t = 0$ and $B_\mathrm{tot}^t = 100$. We then, with a nested Monte Carlo simulation procedure described below, develop a 95$^\mathrm{th}$ percentile upper limit, $A_{95}$, for the data realization. At a given $N_\mathrm{bins}$, we repeat this procedure 400 times, such that we have an ensemble of $A_{95}$ realizations, from which we can compute a sample median and containment intervals. We repeat this entire procedure at each value of $N_\mathrm{bins}$ considered for the asymptotic sensitivity estimates.

\subsubsection{Limit-Setting via the Neyman Procedure}
For a given data realization, we develop a 95$^\mathrm{th}$ percentile upper limit via Neyman's construction of confidence intervals via a Monte Carlo procedure. To do so, we first define the test-statistic for upper limits of \cite{Cowan:2010js}, which is given by
\begin{equation}
q(A|{\vec d} )=
\begin{cases}
-2 \ln \displaystyle\frac{\mathcal{L}\!\left({\vec d} \mid A,\hat{B}_\mathrm{tot}(A , {\vec d})\right)}
{\mathcal{L}\!\left( {\vec d} \mid \hat{A}({\vec d}) ,\hat{B}_\mathrm{tot}({\vec d}) \right)} & \text{if } A>\hat{A},\\
0 & \text{otherwise,}
\end{cases}
\end{equation}
with 
\begin{equation}
\begin{gathered}
\hat{B}_\mathrm{tot}(A, {\vec d} )=\arg\max_{B_\mathrm{tot}}\,\mathcal{L}({\vec d} \mid A,B_\mathrm{tot}) \\
 \hat{A}(\bm{d}),\, \hat{B}_\mathrm{tot}({\vec d})=\arg\max_{A,B_\mathrm{tot}}\,\mathcal{L}({\vec d} \mid A,B_\mathrm{tot}).
\end{gathered}
\end{equation}
To generate a limit $A_{95}$ given a dataset ${\vec d}$, we evaluate $q(A \mid {\vec d})$ and $\hat{B}_\mathrm{tot}(A)$. The $95^\mathrm{th}$ percentile upper limit is then defined as the value of $A_{95}$ such that $q(A_\mathrm{95} |{\vec d})$ is at the $95^\mathrm{th}$ percentile of the ensemble in the distribution of $q(A_\mathrm{95} |{\vec d}^A)$, where data ${\vec d}^A$ are drawn from the model with parameters  $A_\mathrm{95}$ and $\hat{B}_\mathrm{tot}(A_{95}, {\vec d})$.

In practice, we determine $A_{95}$ by considering a candidate value of $A$, generating $16,384$ realizations of data ${\vec d}^A$ under the model specified by $A$ and $\hat{B}_\mathrm{tot}(A, {\vec d})$, from which we can develop $16,384$ samples of $q(A | {\vec d}^A)$ and make a noisy estimate of the quantile of $q(A |{\vec d})$ in the distribution of $q(A | {\vec d}^A)$. We repeat this procedure for a range of roughly 10 values of $A$ which realize samples for which $q(A | {\vec d})$ is at a quantile between roughly $0.94$ and $0.96$, representing a range which is much broader than the statistical error on the quantile estimate based on the 16,384 samples. After this repeated procedure, we have sample quantiles $\hat{Q}(A_i)$ for each of the candidate signal strengths $A_i$. We fit a quadratic function, assuming equal errors to the $\{A_i, \hat{Q}(A_i)\}$ data, to develop a prediction for the quantile $Q(A)$ as a function of $A$, determining $A_{95}$ by $Q(A_{95}) = 0.95$. This procedure ensures that the sample error in setting the $95^\mathrm{th}$ percentile limit on a given dataset is negligibly small for the purposes the comparison here.

\subsection{Results}

We present the results of our asymptotic projections with Monte Carlo expectations for the limit on the signal strength parameter $A$ under the null in the right panel of Fig.~\ref{fig:monte_carlo_demonstration}. The datapoints, which are plotted as a function of $N_\mathrm{bins}$, indicate the median $A_{95}$, with the three errorbars indicating, from top to bottom, the 97.5$^\mathrm{th}$ percentile, $84^\mathrm{th}$ and $16^\mathrm{th}$ percentile values of $A_{95}$ evaluated from the Monte Carlo ensemble of 400 realizations of $A_{95}$ determined by the Neyman construction. By comparison, dashed black line indicates the asymptotic expectation for the median $A_{95}$, while the green and yellow bands indicate the  $16^\mathrm{th}$, $84^\mathrm{th}$, and 97.5$^\mathrm{th}$ percentile values of $A_{95}$ in the asymptotic limit.

Overall, the agreement between the asymptotic expectations and the Monte Carlo results is very good, and notably persists across the full range of $N_\mathrm{bins}$. For the smallest value $N_\mathrm{bins} = 2$, the expectation of 50 counts per bin is reasonably well-approximated as Gaussian, while for the largest value, $N_\mathrm{bins} = 500$, the expected number of counts per bin is $0.2$, which is deep in the Poissonian regime. The $\sim10\%$ discrepancies are likely primarily driven by $O(N^{1/2})$ deviations from the asymptotic expectations in the finite-sample regime \cite{Cowan:2010js}, where $N$ is the total number of events. This means that data collections with fewer than $\sim 100$ counts may require a more careful consideration. However, in broad generality, we find these Monte Carlo tests to provide strong validation of our previous Fisher forecasts.

\section{Shifted Sidereal Stacking}
\label{sec:Sidereal}

\begin{figure}
\centering
\includegraphics[width=0.49\textwidth]{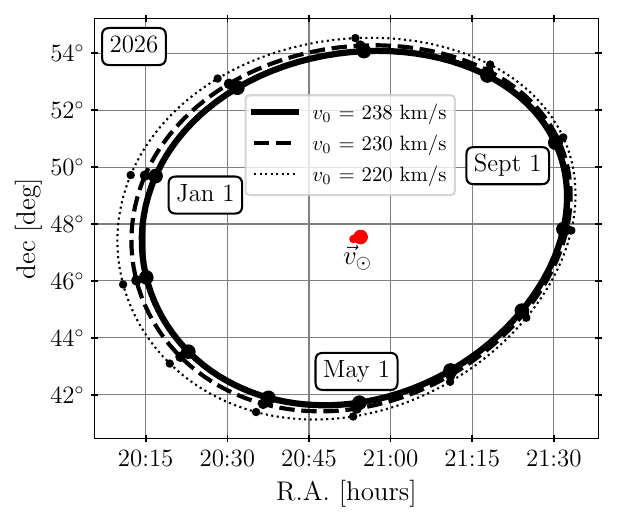}
\caption{The right ascension (RA) and declination (dec) of the Earth velocity vector in the galactic rest frame as a function of time of year, with the RA angle given in hours ($360^\circ = 24$~hours).
The speed of the local standard of rest $v_0$ is not known precisely: following~\cite{Baxter:2021pqo} we adopt $v_0 = 238$\,km/s as the standard value. The dashed and dotted lines show $v_\oplus(t)$ assuming $v_0 = 230$\,km/s and 220\,km/s, respectively.
For reference, the red dots in the center show the constant Sun velocity $v_\odot$ for each value of $v_0$. The first day of each month is marked on the plot, starting with 2026~Jan.~1 (00:00:00 UTC). Note that the angle between the north pole and the Earth velocity is $(90^\circ - \text{dec})$. 
}
\label{fig:RAdec}
\end{figure}

Due to the motion of the Earth around the Sun, the apparent source of the dark matter wind changes its position on the sky by a few degrees every month. Compared to the annual variation in $|\vec v_\oplus(t)|$, the time dependence of $\hat v_\oplus$ has received much less attention. It has a noticeable impact on the expected $R(t)$ signal: from September to December, the expected ``rising'' and ``setting'' times of the DM wind shift by more than 75 minutes. 
As a result, the scattering rate $R(t)$ is not exactly periodic over the sidereal day: both its amplitude and phase should exhibit annual variation. Figure~\ref{fig:RAdec} quantifies this change, showing the sky position of $\vec v_\oplus(t)$ in astronomer coordinates of declination (angle above the celestial equator) and right ascension (azimuthal angle). 

The statistical procedures outlined in the previous section work perfectly well for an aperiodic $R(t)$, as long as the data from each sidereal day is kept separate. 
Stacking the data over the sidereal day would tend to smear out the signal, modestly reducing its statistical significance.
Happily, this problem has a relatively simple solution. Rather than binning the data according to sidereal time-of-day, one can bin it according to the time since the most recent ``DM noon,'' defined as the moment when the DM wind was closest to the zenith. 
Expressed in terms of this shifted sidereal time $\tilde{t}$, the phase of $R(\tilde t)$ remains almost exactly constant over the course of the year: the phase changes by only a few seconds, rather than 75 minutes. 
With shifted sidereal time, multiple weeks or months worth of data can be stacked into a single 24\,h span of $\tilde t$, without any destructive interference in the theory predictions for $R(\tilde t)$.

\begin{figure}
\centering
\includegraphics[width=0.49\textwidth]{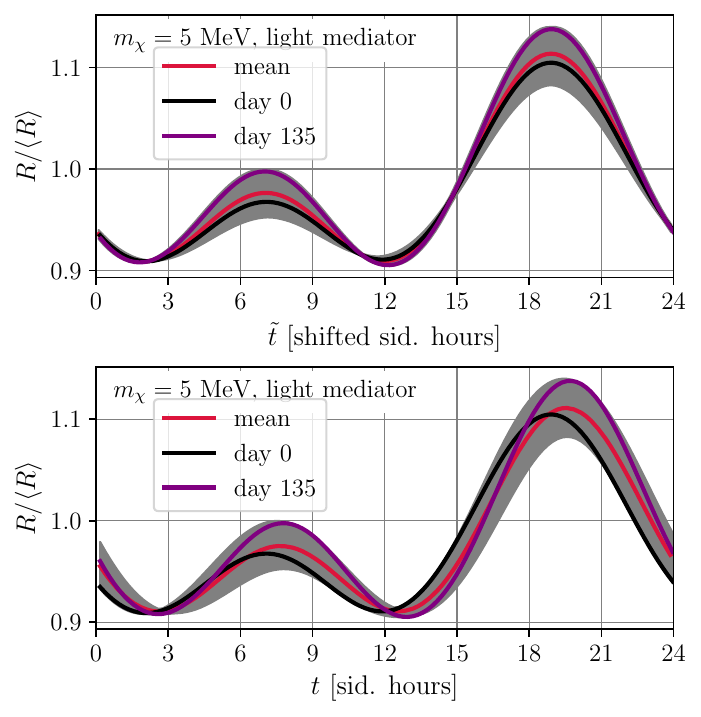}
\caption{Two schemes for stacking daily modulation data are shown, for 366 sidereal days worth of exposure time. The lower panel tracks $R(t)$ according to the sidereal time, while the upper panel uses the shifted sidereal time $\tilde{t}$. The rates on day zero (Jan 1, 2026) and day 135 (May 16) are highlighted in black and purple, with the average value of $R(t)$ shown in red. On the other days of the year, $R(t)$ varies within the gray envelope.}
\label{fig:sidshift}
\end{figure}

Figure~\ref{fig:sidshift} compares the annual variation in the sidereal time-of-day rate $R(t)$ to that of the shifted sidereal time, $R(\tilde t)$. Over the course of a year, the right ascension and declination of the DM wind change by several degrees, shifting the daily $R(t)$ in time and changing its overall shape. The lower panel of Fig.~\ref{fig:sidshift} shows how $R(t)$ responds to the combination of effects: the value of $R(t)$ varies within a window of $0.05\, \langle R \rangle$ at most values of the sidereal time $t$. 
Considering that the average $R(t)$ varies by only $\pm 0.1 \, \langle R\rangle$ about its mean value, the variation in the DM wind direction causes a substantial change in the expected modulation signal. 

The shifted sidereal time $\tilde t$ adjusts for the varying right ascension of the DM wind, producing the $R(\tilde t)$ shown in the top panel of Fig.~\ref{fig:sidshift}. This $R(\tilde t)$ varies within a substantially smaller envelope, especially where $R(t)$ changes most quickly, and the maxima and minima of $R(\tilde t)$ are aligned in $\tilde t$. The variations in the overall amplitude of $R$ caused by the changing declination angle remain; however, the data can now be stacked in $\tilde t$ without destructive interference. 

These stacking procedures are useful for isolating the sidereal daily modulation from e.g.,~annual effects, or any other variations in the rate that occur on non-sidereal timescales. They also provide a lower-dimensional space for statistical analyses, which can reduce the computational expense of multivariate Monte Carlo simulations.
Section~\ref{sec:stackexample} provides an example, where a modest sidereal DM signal is extracted from a larger 24\,h modulating background.

\section{Example Analyses}
\label{sec:Example}

\begin{table}
\centering
\begin{tabular}{| l | l | } \hline
\textbf{Background Type:} & \textbf{Strategy:} \Tstrut\Bstrut\\ \hline
    \,\begin{tabular}{l}
~constant, uncorrelated
\end{tabular}
& 
    \,\begin{tabular}{l}
maximize $f_\text{RMS}$ in \\  every detector:      \eqref{eq:frms} 
    \end{tabular}
\Tstrut\Bstrut\\ \hline 
    \,\begin{tabular}{l}
        uncorrelated and/or \\ correlated, known \\ frequency, unknown phase 
    \end{tabular}
& 
    \,\begin{tabular}{l}
        employ multiple detectors  in \\ different orientations: \eqref{eq:multicorrel}
    \end{tabular}
\\ \hline
\end{tabular}
\caption{Summary of the types of background and the corresponding analysis strategies considered in this work. We generally assume that the total background rate is not precisely known.}
\end{table}

Now equipped with the necessary formalism to perform realistic analyses with synthetic datasets, we proceed to some example analyses with simplified toy models that approximate experimentally motivated detector configurations and backgrounds. 

We adopt a general model for summed signal and background rates measured by a given detector given by~\cite{Kahn:2021ttr}
\begin{equation} \label{eq:bkgprofile}
    R(t) = A \, s(t | \theta, \phi) + B_\mathrm{osc.} \cos(2 \pi f_\mathrm{osc.} t + \phi_\mathrm{osc.}) + B_\mathrm{flat}
\end{equation}
where  $A = R_\chi^{\rm iso}$ is the normalization for the signal rate obtained from the isotropic (angular-averaged) scattering rate
\begin{equation}
   R_\chi^{\rm iso} = \frac{M_{T}}{\rho_T}\frac{\rho_\chi}{m_\chi}\int\! \frac{q dq}{(2\pi)^2}d\omega \, \eta(v_{\rm min(q,\omega)}) \frac{\pi \bar{\sigma}_{\chi T}(q)}{\mu_\chi^2}S(q, \omega). 
   \label{eq:Riso}
\end{equation}
Here, $M_{T}$ is the detector mass; $\rho_T$ is the target density; $\rho_\chi$ and $m_\chi$ are the DM density and mass, respectively; $q$ and $\omega$ are the momentum and energy transfer in the dark matter scattering event, respectively; $\eta(v_{\rm min}) = \int_{v_{\rm min}}^\infty d^3 \mathbf{v}\, \frac{f_\chi(\mathbf{v})}{v}$ is the mean inverse DM speed;
$v_\text{min}(q, E) = E/q + q/(2 m_\chi)$; $\bar{\sigma}_{\chi T}(q)$ is a fiducial momentum-dependent cross section between DM and detector target $T$; and $S(q, \omega)$ is the dynamic structure factor of the target material, which for simplicity we take to only include transitions to the first excited state of trans-stilbene~\cite{Blanco:2021hlm}, with an energy gap of $\omega \approx 4.2$\,eV above the ground state. For illustrative purposes, we take the detector parameters corresponding to $M_T = 1 \ {\rm kg}$ of trans-stilbene ($\rho_T = 1.14 \ {\rm g}/{\rm cm}^3$), $
\rho_\chi = 0.4 \ {\rm GeV}/{\rm cm}^3$, $\eta(v_{\rm min})$ corresponding to the Standard Halo Model with Earth velocity $v_\oplus = 250$\,km/s relative to the Milky Way, and a dark matter mass of 5 MeV. 

We emphasize that the detector optimization can in principle depend strongly on the prior for the DM signal model. This is already implicit in the use of $S(q,\omega)$ in Eq.~(\ref{eq:Riso}), which is the appropriate detector response function for DM which couples to electron density, but not necessarily for other spin- and velocity-independent operators~\cite{Kahn:2021ttr,Krnjaic:2024bdd,Catena:2024rym,Hochberg:2025rjs,Giffin:2025hdx}. To illustrate this point, in Sec.~\ref{sec:SingleDetectorExamples} below, we will consider the cases of both heavy and light mediators, corresponding to
\begin{equation}
 \bar{\sigma}_{\chi T}(q) = \begin{cases}
 \bar{\sigma}_{\chi e} &, \qquad {\rm heavy \ mediator} \\
\bar{\sigma}_{\chi e}(q_0/q)^4 &, \qquad {\rm light \ mediator}
 \end{cases}
\end{equation} 
respectively, where the reference momentum scale for the light mediator is $q_0 = \alpha m_e \approx 3.7 \ {\rm keV}$. In Sec.~\ref{sec:MultiDetectorOptimization}, we restrict exclusively to the case of a light mediator in order to focus on the effect of changing background models. For illustrative purposes in both analyses, we take $\bar{\sigma}_e = 10^{-38} \ {\rm cm}^2$, which is just below the current sensitivity of DAMIC-M to the kinetically-mixed light dark photon scenario at $m_\chi = 5 \ {\rm MeV}$~\cite{DAMIC-M:2025luv}. 

To calculate the anisotropic scattering rate, we take the molecular form factor $f_s^2(\vec q) \propto S(\vec q, \omega)$ for trans-stilbene from Ref.~\cite{Blanco:2021hlm}. We use the code Vector Spaces for Dark Matter (VSDM)~\cite{Lillard:2023qlx,Lillard:2023cyy,Lillard:2025aim} to evaluate the scattering rate for 16,200 detector orientations $\mathcal R \in SO(3)$ for each DM model. For the Standard Halo Model, this is a sufficiently fine sampling of $SO(3)$ that the scattering rate for any other rotation operator $\mathcal R$ can be found by interpolation.

For the background, $B_\mathrm{osc.}$, $f_\mathrm{osc.}$, and $\phi_\mathrm{osc.}$ are the amplitude, oscillation frequency, and phase of a sinusoidally modulating background rate, and $B_\mathrm{flat}$ is the amplitude of the constant background rate. The time-dependent dark matter signal profile $s(t)$ is in turn specified by the detector orientation parameters $\theta_n$ and $\phi_n$, which enter the (possibly anisotropic) dynamic structure factor $S(\mathbf{q}, \omega)$, and the anisotropic halo integral~\cite{Coskuner:2019odd,Kahn:2021ttr,Lillard:2023cyy}
\begin{equation}
    \eta(\mathbf{q}, \omega; t) = 2q \int\! d^3 \mathbf{v} \, f(\mathbf{v} + \mathbf{v}_\oplus(t)) \,\delta\!\left(\omega - \mathbf{q}\cdot \mathbf{v} - \frac{q^2}{2m_\chi}\right)
\end{equation}
through the relative orientation of the detector and the DM wind. As in previous sections, integrating the instantaneous event rate $r(t)$ over the finite-duration time bins yields the model prediction for the expected counts in each bin. We compactly denote this model prediction as
\begin{equation*}
    \bm{\mu}(\bm{\Theta}, {\bm \theta}, {\bm \phi})
\end{equation*}
where $\bm{\Theta} = \{A,B_\mathrm{osc.}, f_\mathrm{osc.}, \phi_\mathrm{osc.}, B_\mathrm{flat}\}$ is the parameter vector for the signal normalization and background profiles, and ${\bm \theta} = (\theta_1, \theta_2, \dots, \theta_{N_{\rm det.}})$ and ${\bm \phi} = (\phi_1, \phi_2, \dots, \phi_{N_{\rm det.}})$ are the orientation angles of each of the $N_{\rm det.}$ detectors. In this parameterization, we have already removed all rotational degeneracy because $\theta_i, \phi_i$ are defined with respect to the DM wind vector at $t = t_{\rm ref.}$ In particular, performing an SO(3) rotation on all detector angles $(\theta_i, \phi_i)$ simultaneously will give a different signal profile, and thus represents an independent set of signal parameters.

\begin{figure}
\centering
\includegraphics[width=0.49\textwidth]{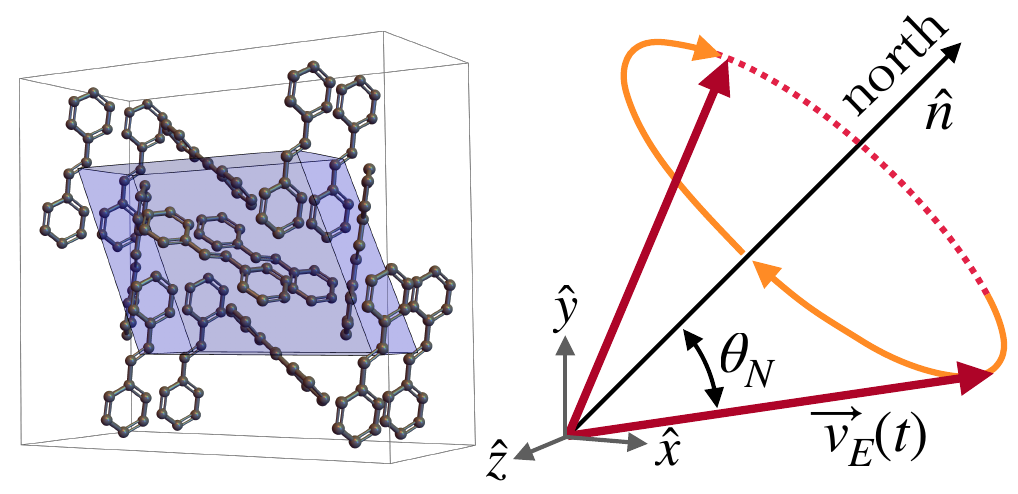}
\caption{The crystal-centric coordinate system. On the left, we show the translationally invariant unit cell of trans-stilbene (outlined in purple). It is symmetric with respect to rotations of $180^\circ$ about the $\hat z$ axis, or inversions $\vec x \rightarrow -\vec x$ through the central point of the molecule. On the right, we show the Earth velocity in the same crystal-centric system, with the sample oriented to place the north pole along the $\hat n = (\theta_n, \phi_n)$ direction. As the Earth rotates, the Earth velocity and DM wind velocity sweep out a cone with opening angle $\theta_N$. }
\label{fig:geometry}
\end{figure}

As in Ref.~\cite{Blanco:2021hlm}, we define a coordinate system based on the unit cell of trans-stilbene, identifying the $\hat z$ direction ($\theta = 0$) with the $\mathbb{Z}_2$ symmetry axis $\hat b$, and the $\hat x$ direction ($\phi = 0$) with the $\hat c$ axis of the crystal. In this section, we parameterize the crystal orientation with two angles, $\theta_n$ and $\phi_n$, where the unit vector $\hat n = (\theta_n, \phi_n)$ indicates the direction of the north pole relative to the crystal unit cell. For example, by aligning the $\hat b = \hat z$ axis with the north pole, we have $\theta_n = 0$, while the $\theta_n = 90^\circ$ orientations put the north pole in the $XY$ plane of the crystal. Every sidereal day, the DM wind $\vec v_\oplus$ sweeps out a cone in $(\theta, \phi)$, a fixed angular distance $\theta_N \equiv 90^\circ - \text{dec}$ from the north pole. 
A diagram of the coordinate system is provided in Fig.~\ref{fig:geometry}.
In this section, we take $\theta_N = 42^\circ$. Figure~\ref{fig:RAdec} shows that this value is appropriate for January~15 or August~1, and it is also close to the annual mean. Throughout the year, $\theta_N$ varies by about $\pm 6^\circ$. 

To fully specify an $SO(3)$ orientation, we specify the rotation about the north pole with a third parameter, $\psi_t$. The only effect of $\psi_t$ is to shift the modulation signal, $R(t) \rightarrow R(t - \psi_t)$, which leaves $\frms$ unchanged (assuming the exposure time is an integer number of sidereal days). While the overall phase is unimportant for the analysis, the relative phases between detectors are highly relevant, particularly in the presence of a modulating background, and we incorporate these relative phases in Sec.~\ref{sec:MultiDetectorOptimization} below. For a single detector, for a given orientation specified by $\theta$ and $\phi$, the likelihood of observing $\mathbf{d}$ under the expected counts $\bm{\mu}( \bm{\Theta}, \theta, \phi)$ can be computed with the standard Poisson likelihood of Eq.~(\ref{eq:PoissonLikelihood}). 

\subsection{Single Detector Optimization and Analysis for Flat Backgrounds}
\label{sec:SingleDetectorExamples}

The simplest possible analysis is one in which the background is known to be constant in time, while the precise rate of background events is unknown. The signal and background models for $R(t)$ are given by \eqref{eq:bkgprofile} with $(A, B_\text{flat})$ and $(A = 0, B_\text{flat})$, respectively. Following Section~\ref{sec:flat}, the optimal detector orientation is found by maximizing $\frms \sqrt{N}$, for $\frms$ given by \eqref{eq:frms}. 

\begin{figure*}
\centering
\includegraphics[width=0.99\textwidth]{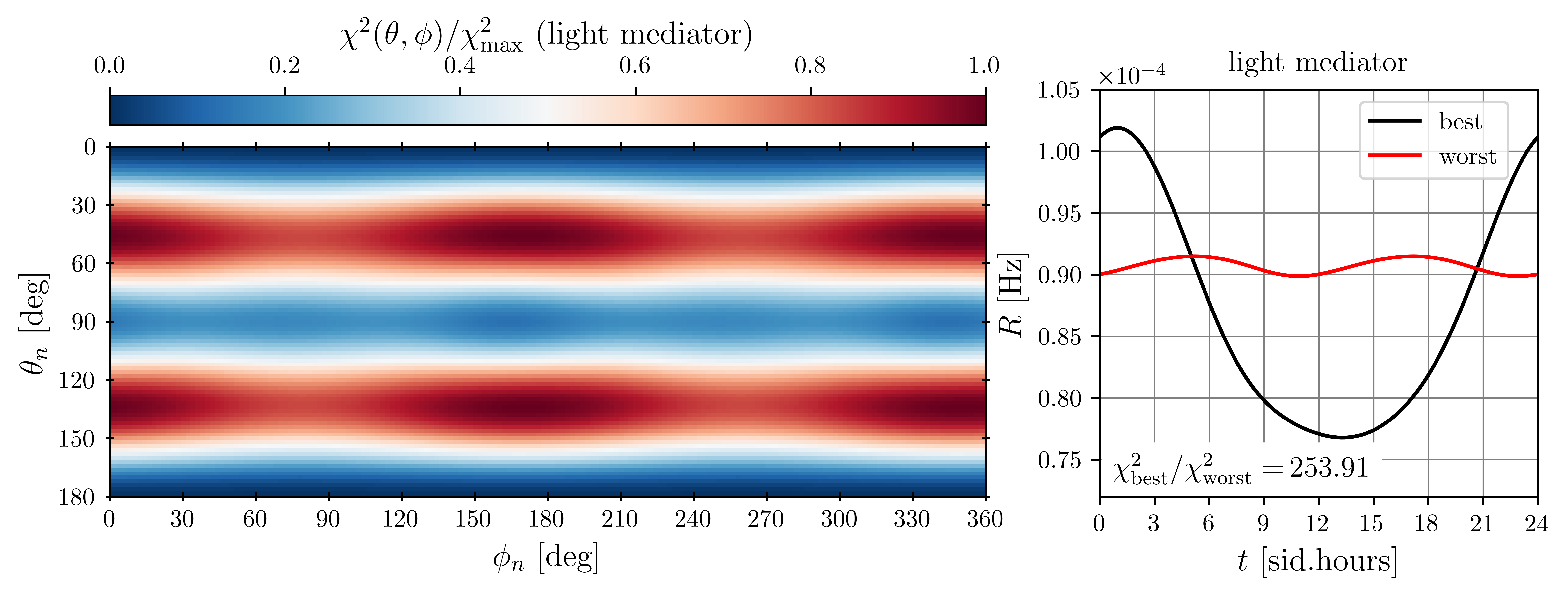}
\includegraphics[width=0.99\textwidth]{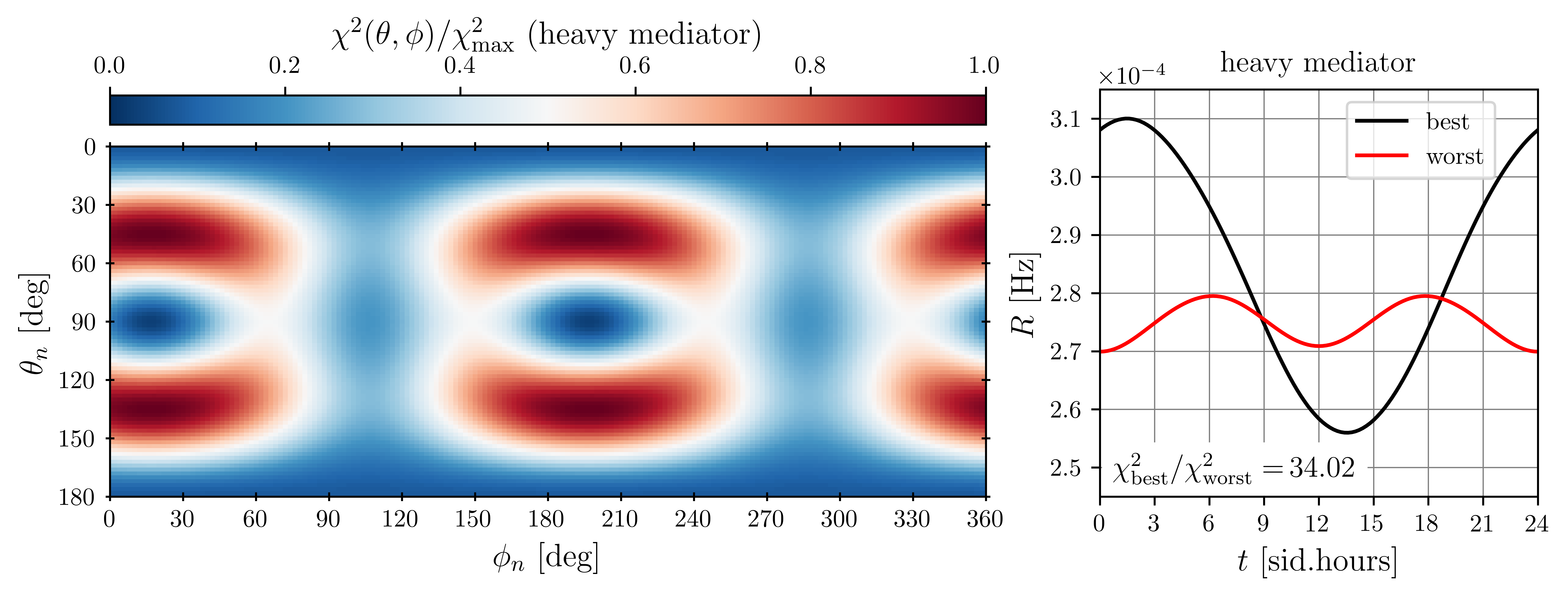}
\caption{Optimized orientations for a single trans-stilbene detector, assuming a constant background, with $m_\chi = 5 \ {\rm MeV}$, $M_T = 1 \ {\rm kg}$, and $\bar{\sigma}_e = 10^{-38} \ {\rm cm}^2$, for both limits of the mediator mass. In the left panel, we plot $\chi^2 = \frms^2 N$ as a function of the detector orientation $(\theta_n, \phi_n)$. To the right, we plot $R(t)$ for the best and worst orientations, corresponding to $(\theta_n, \phi_n) = (46^\circ, 16^\circ)$ and $(90^\circ, 16^\circ)$ in black and red, respectively, for the heavy mediator. For the light mediator, the best and worst scores are found (respectively) at $(\theta_n, \phi_n) = (46^\circ, 170^\circ)$ and $\theta_n = 0^\circ$. 
The respective $\frms$ scores of the best and worst $R(t)$ functions are $6.72\%$ and $1.18\%$ for the heavy mediator, and $10.26\%$ and $0.63\%$ for the light mediator. 
}
\label{fig:flatoptimal}
\end{figure*}

For this example, we determine the significance of a single trans-stilbene detector for both heavy and light mediators as a function of the detector orientation, using the formalism of Sec.~\ref{sec:chi-square}. The results are shown in Fig.~\ref{fig:flatoptimal} (left). The top (bottom) panel corresponds to the light (heavy) mediator signal model; as anticipated, the optimal orientations differ depending on the assumed signal model. In the least-optimal orientations, the $\chi^2$ value is   roughly $\sim 1/30$ of the maximum $\chi^2$ for the heavy mediator, and $\sim 1/250$ of the maximum $\chi^2$ for the light mediator. The ratio will of course vary with the DM mass, but this type of analysis provides a blueprint for experiments to perform their own optimization for a desired DM signal model. Note that the $\chi^2(\theta_n, \phi_n)$ values reflect the symmetries of the detector crystal: the trans-stilbene unit cell is symmetric under $180^\circ$ rotations about the $\hat b$ axis, as well as under central inversion $(\vec x \rightarrow - \vec x)$. For the Standard Halo Model, which is invariant under rotations about $\vec v_\oplus$, the solutions for $(\theta_n, \phi_n)$ will inherit the $\mathbb{Z}_2 \times \mathbb{Z}_2$ symmetry of the unit cell.

The right panel of Figure~\ref{fig:flatoptimal} shows $R(t)$ for the best and worst orientations for this $m_\chi = 5$\,MeV DM candidate. By rotating the detector from the best to the worst orientation, the $\frms$ of the signal decreases by more than a factor of 16 for the heavy mediator, or seven (7) for the light mediator. 
This ability to turn on or off the modulating part of the DM signal by adjusting the detector orientation is a highly useful consequence of directionality.

Notice that the worst orientation in each case yields an $R(t)$ that modulates with an approximate 12 hour period. The amplitude is small, but the DM signal is qualitatively different. As we will show in the next section, these 12\,h periodic DM signals are highly useful whenever an experiment encounters a large 24\,h modulating background.

\subsection{Multi-Detector Optimization and Analysis for Time-Varying Background Models}
\label{sec:MultiDetectorOptimization}

To  proceed beyond the case of single detector which experiences a constant background event rate, we consider  data collection using three independently orientable  detectors over the duration of a single day. We consider three background scenarios:
\begin{enumerate}
    \item \textbf{Non-modulated backgrounds}:  the data are realized under a true model where $B_\mathrm{osc.} = 0.$, such that the true background rate is constant, but we do not assume this in the analysis and allow for a background phase (along with the amplitude) as nuisance parameters.
    \item \textbf{Weakly modulated backgrounds}: the data are realized under a true model where $B_\mathrm{osc.} / B_\mathrm{flat} =  0.1$, resulting in a total true background rate that experiences small but appreciable variation over the modulation period.
    \item \textbf{Strongly modulated backgrounds}: the data are realized under a true model where $B_\mathrm{osc.} / B_\mathrm{flat} =  0.9$, resulting in a total true background rate that experiences significant variation over the modulation period.
\end{enumerate}
For definiteness, we take $B_\mathrm{flat} = 1 \, {\rm Hz}$ in all the examples which follow, and we assume that the true flat background rate is common to all detectors, which would correspond to the physical situation of e.g.~an exogenous radioactive background. As noted in Sec.~\ref{sec:UncorrelatedBkg}, our formalism applies equally well for endogenous backgrounds which vary from detector to detector, for example the carbon-14 rate.

Because environmental effects may generate a time-varying event rate that challenges efforts to detect a time-varying dark matter signal, consideration of how the detector orientation may be chosen to optimize the expected sensitivity is of particular interest here. Because the signal is expected to realize a daily modulation, we will fix $f_\mathrm{osc.} =(24\,\mathrm{hrs})^{-1}$ and assume it to be perfectly known throughout this section. However, consideration of  higher harmonics or more general frequency uncertainty might be of interest for realistic experimental scenarios.
Over a single day, the solar day and the sidereal day ($t_\text{sid} \simeq 0.9973 \times 24\,\text{h}$) are nearly indistinguishable. For this reason, the successful detector designs proposed in this section can even extract a DM signal from a sidereal daily modulating background ($f_\text{osc} = t_\text{sid}^{-1}$). 

In our generative model, we assume that each detector experiences identical time-varying and time-independent backgrounds. However, in our inference process, we only take the modulated background parameters to be identical across detectors, allowing the flat background rate to be independently inferred at each detector.  Regardless of how the data are generated (whether from the strongly modulated, weakly modulated, or non-modulated  background model), we always allow our hypothesized model to include a modulated background. This represents the \textit{a priori} uncertainty as to whether or not a modulated background is present which can only be interrogated after data collection has begun. That is, our likelihood for the total dataset $\mathbf{d} = \{\mathbf{d}^{(1)}, \mathbf{d}^{(2)}, \mathbf{d}^{(3)}\}$ comprised of the data from each detector is given by
\begin{equation}
\begin{split}
\mathcal{L}&(\mathbf{d} | A, \bm{\Theta}, \bm{\theta}, \bm{\phi}, \bm{\psi}_t) \\
&= \prod_i \mathcal{L}(\mathbf{d}^{(i)} | A, \{B_\mathrm{osc.}, f_\mathrm{osc}, \phi_\mathrm{osc}, B_\mathrm{flat}^{(i)}\}, \, \theta^{(i)}, \phi^{(i)}, \psi_t^{(i)})
\label{eq:joint_likelihood}
\end{split}
\end{equation}
where the total background parameter vector is now $\bm{\Theta} = \{B_\mathrm{osc.}, f_\mathrm{osc}, \phi_\mathrm{osc},  B_\mathrm{flat}^{(1)}, B_\mathrm{flat}^{(2)}, B_\mathrm{flat}^{(3)}\}.$
and we have taken the product of the Poisson likelihoods for the data collected by each detector in Eq.~(\ref{eq:joint_likelihood}). For compactness of notation we have collected the three phase offsets $\psi_t^{(i)}$ for each detector into the signal parameter vector $\bm{\psi}_t$, but there is an overall phase degeneracy corresponding to the common phase between all three detectors that we will remove in our analysis by setting the phase of detector 1 to a reference value and taking ${\bm \psi}_t \equiv (\psi_t^{(2)} - \psi_t^{(1)}, \psi_t^{(3)} - \psi_t^{(1)})$.

\subsubsection{Analysis and optimization with known background phases}

A relatively simple scenario is represented by the one in which the background phase $\phi_\mathrm{osc.}$ is known while the background rates $B_\mathrm{osc.}$ and $B_\mathrm{flat}$ are unknown. Such a scenario might be realized if periodic environmental effects have been independently measured but the precise magnitude  of the backgrounds event rate they generate is unknown. We take $\phi_\mathrm{osc.}$ in both our generative model and hypothesized model to be $0$, but the results we present are robust with respect to the choice of fixed $\phi_\mathrm{osc.}$ because alternate choices of $\phi_\mathrm{osc.}$ can be accommodated by a suitable transformation of the $\theta, \phi$ which specify the detector orientation.  

Taking the background amplitude $B_\mathrm{osc.}$ and $B_\mathrm{flat}$ to be the unknown nuisance parameters, we can estimate our expected sensitivity to the signal amplitude parameter $A$ using Eq.~(\ref{eq:FisherVariance}). Because this sensitivity depends on the detector orientations, we denote it $\sigma_A(\bm\theta, \bm\phi, \bm \psi_t)$, and we can determine the optimal detector orientations by
\begin{equation}
    \bm\theta^\mathrm{opt.}, \bm\phi^\mathrm{opt.}, \bm\psi_t^\mathrm{opt.} = \mathrm{argmin}_{\bm{\theta}, \bm{\phi}, \bm{\psi}_t} \left[\sigma_A(\bm{\theta}, \bm{\phi}, \bm{\psi}_t)\right],
\end{equation}
with the optimal sensitivity given by
\begin{equation}
    \sigma_A^\mathrm{opt.} = \sigma_A(\bm\theta^\mathrm{opt.}, \bm\phi^\mathrm{opt.}, \bm\psi_t^\mathrm{opt.}).
\end{equation}
Note that in this single detector case, the optimization in detector orientation could be straightforwardly accomplished by maximizing the $f_\mathrm{RMS}$ quantity defined by Eq.~(\ref{eq:fRMS_chi2}) as a function of $(\theta, \phi)$ along the lines demonstrated in Sec.~\ref{sec:SingleDetectorExamples}, with the overall phase $\psi_t$ irrelevant to the optimization. Now, however, since the detectors may be each independently oriented, our detector orientation optimization is now requires a considerably more complex optimization in the eight-dimensional space of $\bm{\theta}$, $\bm{\phi}$, and the two relative phases $\psi_t^{(2)} - \psi_t^{(1)}$ and $\psi_t^{(3)} - \psi_t^{(1)}$. In general, this optimization can only performed numerically, and care must be taken as the many symmetries of the problem lead to numerous degenerate or near-degenerate sensitivity maxima. In this work, we handle this challenge through optimizations performed with the differential evolution global optimizer \cite{Storn1997} and do not attempt to distinguish between or identify the potentially many equally optimal detector orientations; the results we present merely reflect one optimal arrangement.

\begin{figure*}[!ht]
\centering
\includegraphics[width=0.99\textwidth]{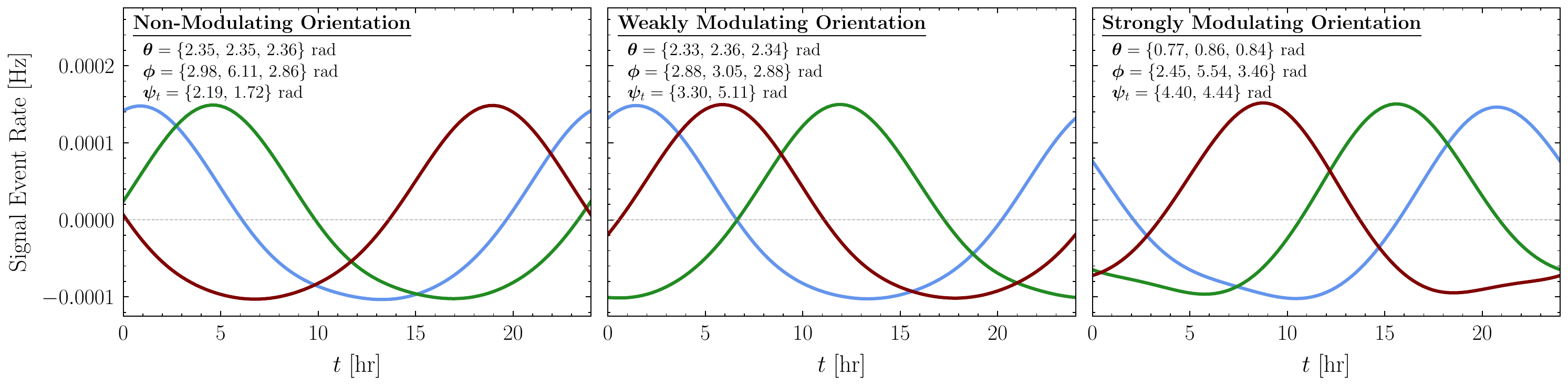}
\caption{(\textit{Left panel}) Mean-subtracted event rates for three identical detectors oriented so as to maximize the sensitivity of data collection in the non-modulating (\textit{i.e.}, flat) background scenario for analyses in which the phase of an inferred modulating background is taken to be fixed and known at $\phi_\mathrm{osc.}=0$. In  the upper left corner, we indicate the $\bm{\theta}$, $\bm{\phi}$, and $\bm \psi_t$ which specify the optimal orientations and phases used to generate the event rates. (\textit{Middle panel}) As in the left panel, but for the set of detector orientations which maximize the sensitivity in the weakly modulating background scenario.  (\textit{Right panel}) As in the left panel, but for the set of detector orientations which maximize the sensitivity in the strongly modulating background scenario.}
\label{fig:known_multiple_detector}
\end{figure*}

In this multi-detector context, we perform this sensitivity maximization for each of our three background scenarios, with results summarized in Fig.~\ref{fig:known_multiple_detector}, which illustrates the mean-subtracted time-dependent signal profiles that optimize the expected sensitivity for the strongly modulated, weakly modulated, and non-modulated background scenarios for the known phase scenario. We find that our optimal sensitivity to the signal amplitude parameter is given by 18.1, 22.1, and 22.1 in the strongly modulated, weakly modulated, and non-modulated scenarios, respectively. By comparison, if all three-detectors were operated in the configuration that optimized the independent sensitivity of a given detector, the optimal sensitivity would be 19.5, 24.1, and 22.1 in the strongly modulated, weakly modulated, and non-modulated scenarios, respectively. 

The sensitivity gain here from a multi-detector optimization is rather marginal; if the phase of the background is initially known, then it relatively easy to choose single orientation of the orientation such that there is very little degeneracy between the signal and the background. However, we will see considerable improvements in projected  sensitivities for scenarios where the background phase is unknown prior to a data collection, which we demonstrate in detail in Sec.~\ref{sec:MultiDetectorUnknownPhases}.

\subsubsection{Analysis and optimization with unknown phases}
\label{sec:MultiDetectorUnknownPhases}

\begin{figure*}[!t]
\centering
\includegraphics[width=0.99\textwidth]{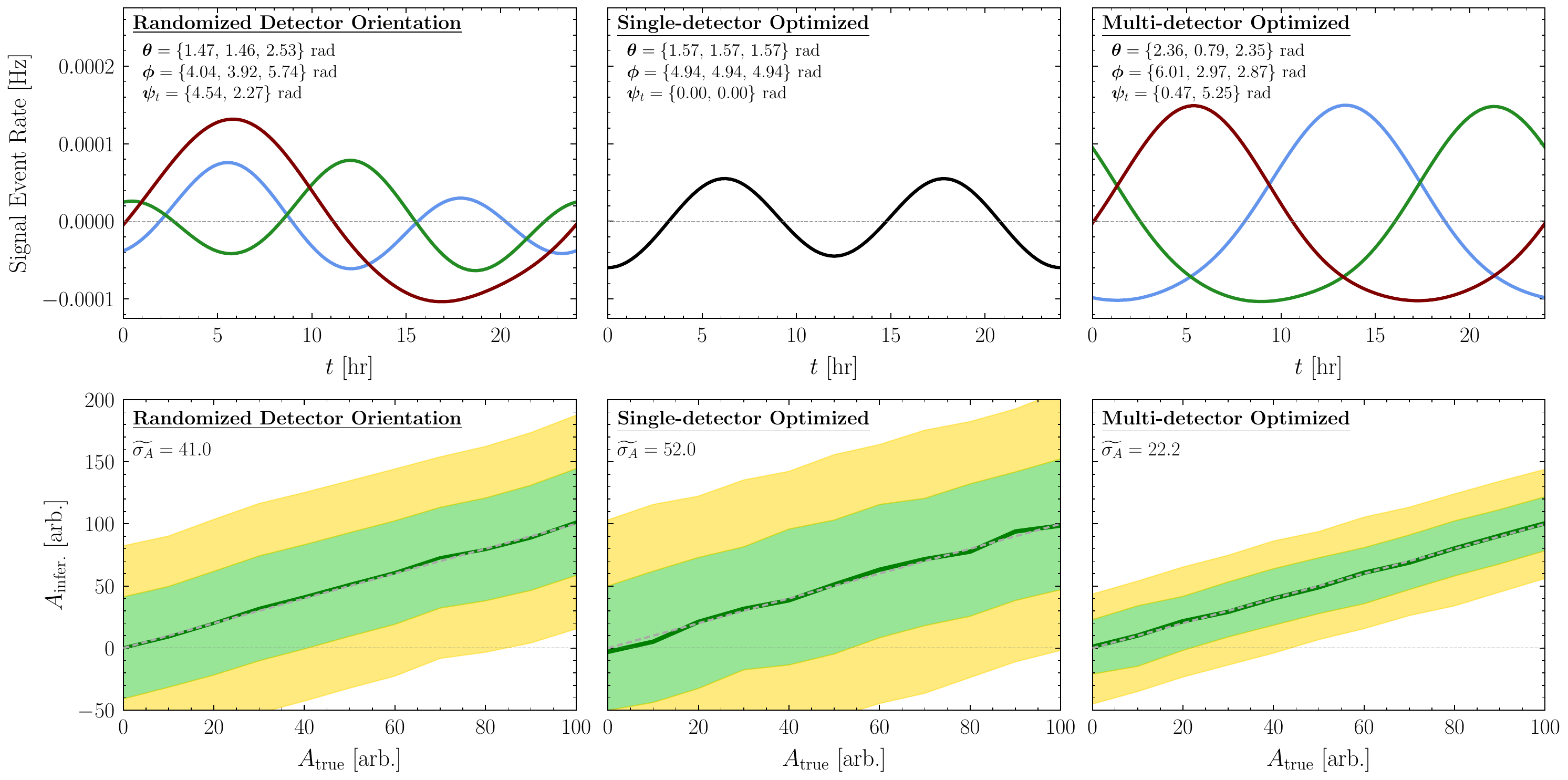}
\caption{Summary figures for the Monte Carlo analysis tests in the non-modulated background. (\textit{Upper left}) Mean-subtracted event rates for three randomly sampled detectors, with $\bm{\theta}$, $\bm{\phi}$, and $\bm \psi_t$ indicated by the text in the upper left. (\textit{Lower left}) In solid green, the mean best-fit signal strength as a function the true signal  strength under which the data were generated. Inference is performed in each pseudo-experiment using a randomized detector orientation. The mean $1\sigma$ confidence interval, which represents the $1\sigma$ upper limit and $1\sigma$ lower limit on the signal averaged over the pseudo-experiments, is shown in green as a function of true signal strength. Similarly, the mean $2\sigma$ confidence interval, which represents the $2\sigma$ upper limit and $2\sigma$ lower limit on the signal averaged over the pseudo-experiments, is shown in gold as a function of true signal strength. The expected sensitivity, corresponding to the difference between the expected best-fit and the expected $1\sigma$  upper (or lower) level, is indicated by text in the upper left corner.  (\textit{Upper center}) As in the upper left panel, but depicting the event rate for each of the three detectors when all are oriented in a manner which maximizes their individual sensitivity. (\textit{Lower center}) As in the lower left panel, but illustrating the sensitivity realized when all of the detectors are oriented such that their individual sensitivity is maximized. That the joint sensitivity is worse here than in the fully randomized scenario is visually clear by the wider $1\sigma$ and $2\sigma$ intervals.  (\textit{Upper right}) As in the upper left panel, but depicting the event rate for each of the three detectors when oriented in a manner which maximizes their joint sensitivity. (\textit{Lower right}) As in the lower left panel, but illustrating the sensitivity realized when all of the detectors are oriented such that their joint sensitivity is maximized. That the joint sensitivity is better here than in the fully randomized scenario is visually clear by the narrower $1\sigma$ and $2\sigma$ intervals.}
\label{fig:non_modulated_unknown_multiple_detectors}
\end{figure*}

\begin{figure*}[!t]
\centering
\includegraphics[width=0.99\textwidth]{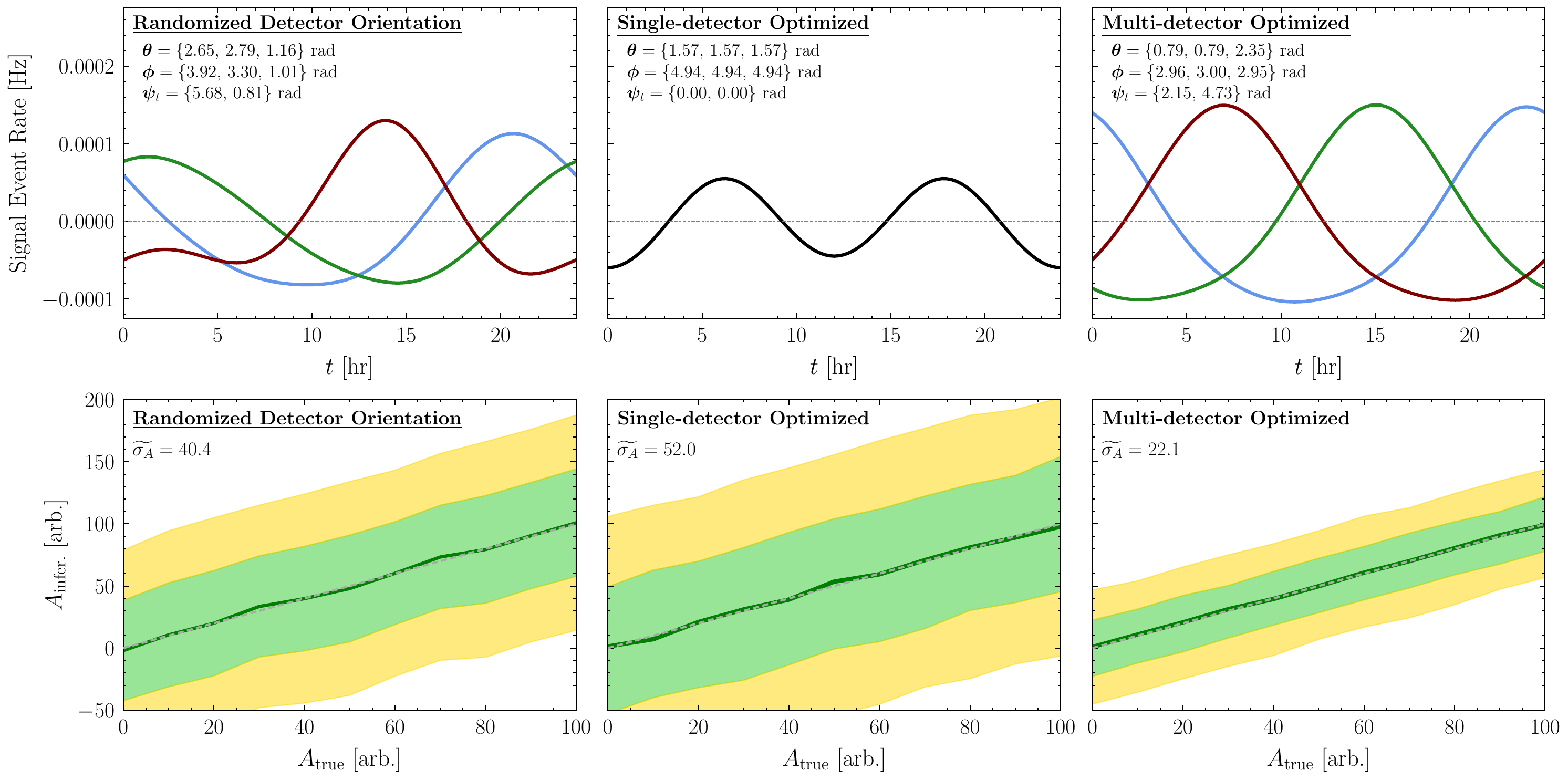}
\caption{As in Fig.~\ref{fig:non_modulated_unknown_multiple_detectors}, but for the weakly modulated background scenario.}
\label{fig:weakly_modulated_unknown_multiple_detectors}
\end{figure*}

\begin{figure*}[!t]
\centering
\includegraphics[width=0.99\textwidth]{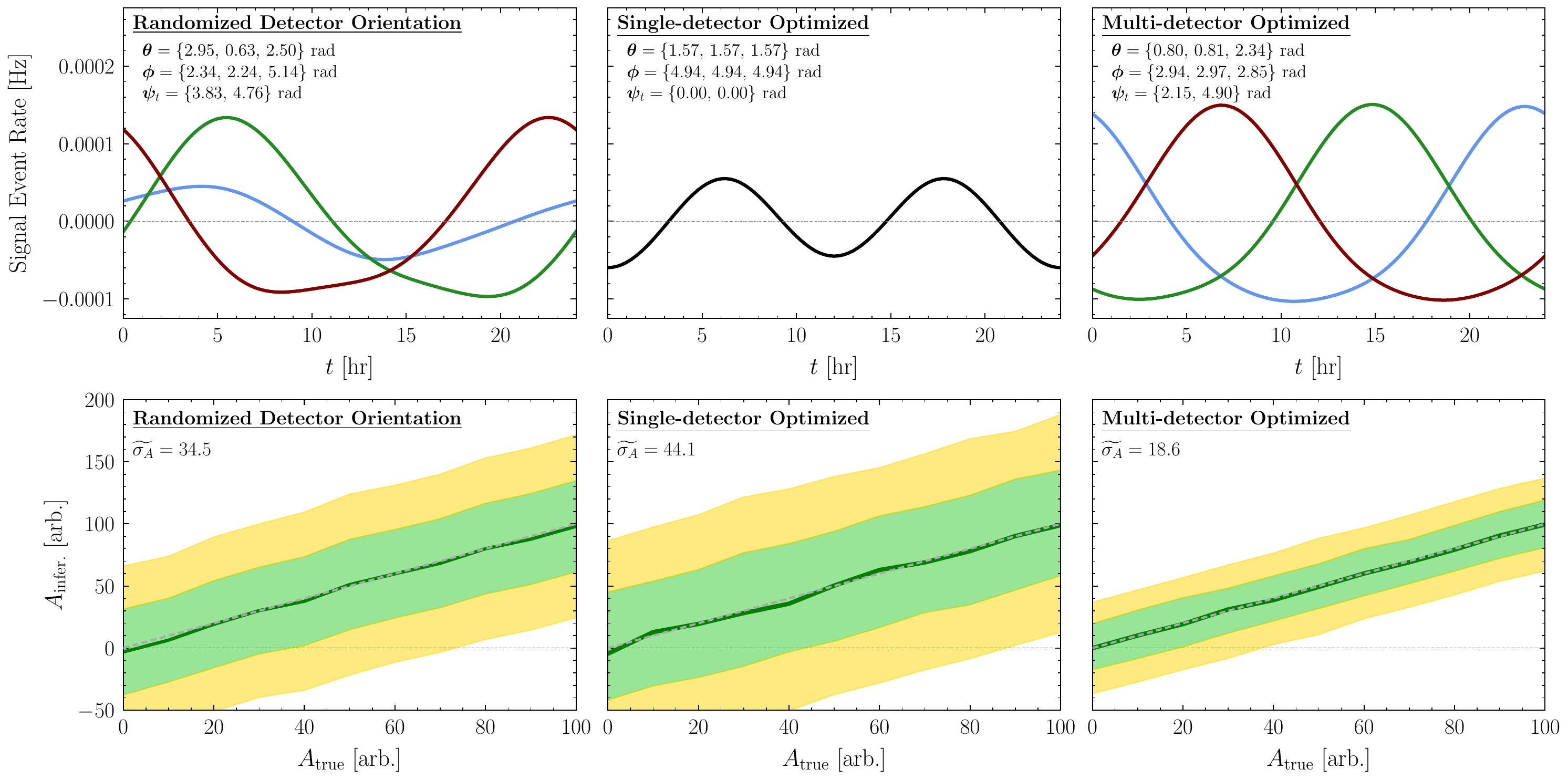}
\caption{As in Fig.~\ref{fig:non_modulated_unknown_multiple_detectors}, but for the strongly modulated background scenario.}
\label{fig:strongly_modulated_unknown_multiple_detectors}
\end{figure*}

Alternatively, experimental considerations might motivate detector configurations which have maximally robust sensitivities even in the presence of modulating backgrounds where even the phases are initially unknown. In this scenario, we promote the $\phi_\mathrm{osc.}$ parameter to a floating nuisance parameter rather than one which is fixed. Moreover, because in this scenario, the phase of the modulating background is unknown \textit{a priori}, rather than seeking to maximize the sensitivity for data generated from a fixed $\phi_\mathrm{osc.}$, we should instead maximize the expected sensitivity to a DM signal after marginalizing over  $\phi_\mathrm{osc.}$ which generates the observed data. To do so, we define the sensitivity to the DM signal inferred for a detector orientation specified by $(\bm\theta,\, \bm\phi, \bm \psi_t)$ when the true data are generated with $\phi_\mathrm{osc.}$ by $\sigma_A(\bm\theta, \bm\phi, \bm \psi_t, \phi_\mathrm{osc.})$,
from which we compute the marginalized (\textit{i.e.}, expected) sensitivity 
\begin{equation}
   \widetilde{ \sigma_{A}}(\bm\theta, \bm\phi, \bm \psi_t) = \int_0^{2\pi} \frac{d\phi_\mathrm{osc.}}{2\pi}\sigma_A(\bm\theta, \bm\phi, \bm \psi_t, \phi_\mathrm{osc.})
   \label{eq:sigmatilde}
\end{equation}
and then the optimal detector configurations and optimal expected sensitivity as 
\begin{equation}
\begin{gathered}
 \bm\theta^\mathrm{opt.}, \bm\phi^\mathrm{opt.}, \bm\psi_t^\mathrm{opt.} = \mathrm{argmin}_{\bm\theta, \bm\phi, \bm \psi_t} \left[\widetilde{\sigma_A}(\bm\theta, \bm\phi, \bm \psi_t)\right] \\
 \widetilde{\sigma_A}^\mathrm{opt.} = \widetilde{\sigma_A}(\bm\theta^\mathrm{opt.}, \bm\phi^\mathrm{opt.},\bm\psi_t^\mathrm{opt.}).
\end{gathered}
\label{eq:optsensitivity}
\end{equation}
As before, we perform this sensitivity maximization for each of our three background scenarios, with results for optimal orientations summarized in the  top panels  of Figs.~\ref{fig:non_modulated_unknown_multiple_detectors},~\ref{fig:weakly_modulated_unknown_multiple_detectors}, and \ref{fig:strongly_modulated_unknown_multiple_detectors}. We find that our optimal sensitivity to the signal amplitude parameter is given by 18.6, 22.1, and 22.2 in the strongly modulated, weakly modulated, and non-modulated scenarios, respectively. By comparison, the expected sensitivity defined by marginalizing the expected sensitivity over the detector orientations and phases $(\theta, \phi, \psi_t)$ is given by 34.5, 40.4, and 41.0. We validate this optimization with a series of Monte Carlo simulations. It is particularly interesting to note that the single-detector optimized configurations achieve generally worse sensitivity, with the marginalized expected sensitivity given by 44.1, 52.0, and 52.0 in the strongly modulated, weakly modulated, and non-modulated scenarios, respectively. 

\begin{figure*}
\centering
\includegraphics[width=0.99\textwidth]{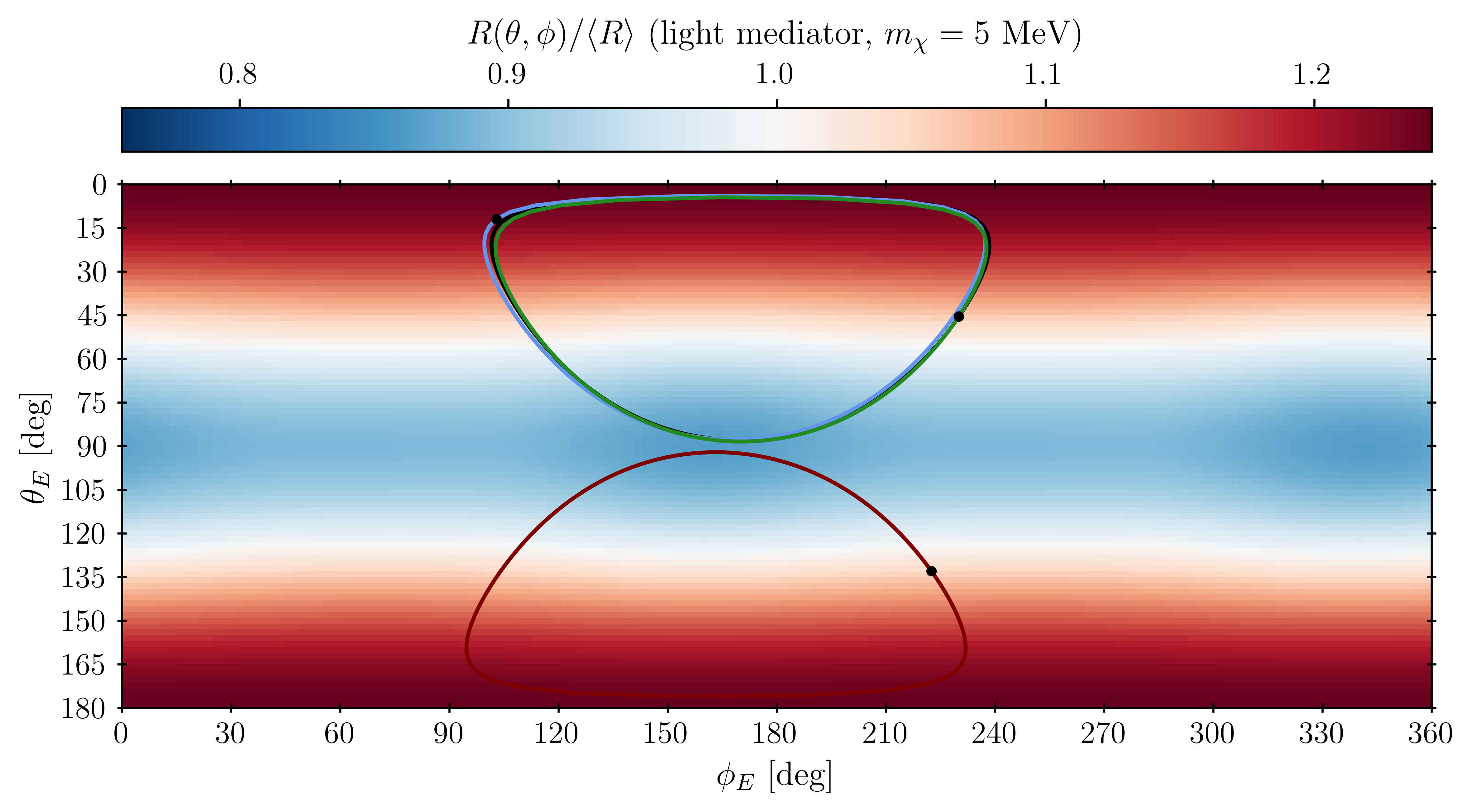}
\caption{The optimal three-detector configuration for the strongly modulated background. The 2d color plot shows $R(\theta_E, \phi_E)$ as a function of the Earth velocity direction $\hat v_\oplus$ in the unit cell coordinate system. In each detector, the Earth's rotation causes $\hat v_E(t)$ to sweep out a cone of opening angle $\theta_N \approx 42^\circ$ centered along the north pole axis. By aligning each detector differently with respect to the north pole, we obtain three distinct but similarly large $R(t)$ signal profiles with different phases, shown in the colored lines. The black dots indicate the  position of each detector at $t = 0$.
In black, we also show the best orientation from Figure~\ref{fig:flatoptimal}, for a single detector with a flat background.
}
\label{fig:orientations}
\end{figure*}

Our Monte Carlo procedure is as follows. Beginning with the non-modulated background scenario, we generate fake data realizations for a given signal strength $A$. We take the detector orientations to be specified (1) either by the optimal orientation determined for the non-modulating background scenario or (2) or by orientations $(\theta^{(i)}, \phi^{(i)})$ which are sampled uniformly on the sphere and relative phases which are distributed uniformly on a sidereal day. We then generate data according to the model specified by $(A, \bm\theta, \bm\phi, \bm\psi_t)$, and the remaining background parameters. After the data are generated, we take all background parameters and the signal strength parameter to be unknown, and we perform a search in data for the presence of a signal. We determine the maximum likelihood estimates of the jointly inferred signal and background parameters by 
\begin{equation}
    \hat{A},\, \hat{\bm{\Theta}} = \mathrm{argmax}_{A, \bm{\Theta}} \mathcal{L}(\mathbf{d} | A, \bm{\Theta}, \bm{\theta}, \bm{\phi}, \bm\psi_t).
\end{equation}
We then determine the $n\sigma$ confidence intervals by the critical values  of hypothesized signal strength such that the test statistic defined by
\begin{equation}
\begin{gathered}
    q(A_{\pm n\sigma}) \equiv -2\log \frac{\mathrm{max}_{\bm\Theta}\mathcal{L}(\mathbf{d}| A_{\pm n\sigma }, \bm{\Theta}, \bm{\theta}, \bm{\phi}, \bm \psi_t)}{\mathcal{L}(\mathbf{d}| \hat{A}, \hat{\bm{\Theta}}, \bm{\theta}, \bm{\phi}, \bm\psi_t)} = n^2 \\
    \mathrm{for} \quad \pm A_{\pm n\sigma } > \hat{A},
\end{gathered}
\end{equation}
in accordance with the asymptotically expected statistics of the test statistic \cite{Cowan:2010js}.

We then perform this procedure many times over a range of $A$, and in repeated pseudo-experiments where data are generated by prescription (2), we independently sample the detector orientations each time. We repeat a similar procedure for data generated in either the strongly modulated or weakly modulated scenario, though in these cases we also independently sample $\phi_\mathrm{osc.}$ uniformly on $[0, 2\pi)$ in each data realization. The simulation and analysis procedures are otherwise identical. Critically, this independent sampling of $\phi_\mathrm{osc.}$ realizes the experimentally relevant scenario of an uninformed uniform prior on the modulation phase.

For each background scenario and each detector orientation choice, we perform 512 Monte Carlo analyses for data generated for 11 true values of $A$ uniformly spaced between $A = 0$ and $A = 50$. From the 512 analysis results at each true value of $A$, we determine the mean values $\hat{A}$, and the $\pm 1\sigma$ and $\pm 2\sigma$ confidence intervals. Figs.~\ref{fig:non_modulated_unknown_multiple_detectors}, ~\ref{fig:weakly_modulated_unknown_multiple_detectors}, and~\ref{fig:strongly_modulated_unknown_multiple_detectors} show the results of the Monte Carlo analyses for our three background scenarios: non-modulated, weakly modulated, and strongly modulated, respectively. In the left panel of each figure, each pseudo-experiment consists of a randomly selected orientation $(\theta_i, \phi_i)$ for each detector on the unit sphere; the signal profile plot at top left shows one such realization. The middle panel of each figure assumes all three detectors are oriented so as to maximize the sensitivity of a single detector; the signal profile is identical to that of the single-detector analysis, but with the detector mass multiplied by a factor of 3. We emphasize that the single-detector optimal configuration is not the one that maximizes, \textit{e.g.}, the peak-to-peak amplitude variation, even in the non-modulated background scenario, because the model used for inference includes a profiled modulating  background that can have appreciable degeneracies with a DM signal. The right panel shows the configuration of the three detectors that minimizes the three-detector analogue of $\widetilde{\sigma_A}$ from Eq.~(\ref{eq:optsensitivity}).

The width of the $1\sigma$ or $2\sigma$ confidence interval in the bottom row of each figure translates directly to a sensitivity for either discovery or exclusion. On average, the $1\sigma$ confidence interval, has a width corresponding to the height of the green interval, while the $2\sigma$ confidence interval has a width corresponding to the height of the yellow interval. Recall that confidence intervals, can be straightforwardly reinterpreted in the context of detection. For example, up to details regarding whether the test is one-sided or two-sided, when the expected $2\sigma$ confidence interval no longer contains the null ($A = 0$), this corresponds to detection at $2\sigma$ significance.

It is clear that the width of the confidence interval is smallest for the optimal detector configuration, improving on the single-detector-optimized configuration by about a factor of 2.4 for all three background scenarios. Because $\sigma_A \propto 1/\sqrt{T}$ from \eqref{eq:significance}, this improvement translates into a 6-fold reduction in the required exposure time to achieve a desired statistical significance. If we assume that the background rate scales with detector mass, then optimizing the detector orientation translates into a commensurate reduction in the required total detector mass for fixed exposure time. Interestingly, in all cases we find that \emph{randomly} orienting the detectors is preferable to optimizing a single detector, which illustrates the statistical power of an anisotropic material to reject both flat and modulating backgrounds.

The specific shape of the signal profile will of course depend on the particular anisotropy of the detector, but the commonalities of the optimized profile among the various background scenarios lend themselves to straightforward interpretation. As we found previously, the optimal profile for a single detector, subject to a background which modulates over a period of a day, is a profile which modulates on a different period; in the case of trans-stilbene, there are orientations which give almost exactly a half-day period. On the other hand, the optimal multi-detector profiles spread out the signal phases throughout the day, consistent with the intuition of Sec.~\ref{sec:correlated} for the two-detector case. Furthermore, the optimal orientation changes fairly little between the weakly- and strongly-modulating cases. Thus, if one expects a modulating background with a 24-hour period, one can achieve close to the optimal sensitivity for a single detector configuration, regardless of the background amplitude. Further improvements are possible with a data-taking procedure that can isolate the background phase, as we discuss in the next section.

Figure~\ref{fig:orientations} presents the three-detector configuration that maximizes the signal significance in the strongly modulated background scenario. Each detector orientation is shown as a cone intersecting the unit sphere, where $(\theta_E, \phi_E)$ are the angular coordinates of the Earth velocity vector $\vec v_\oplus$, and the initial direction $\hat v_\oplus(t = 0)$ is shown by the black dot. In black, we show the best orientation from Fig.~\ref{fig:flatoptimal}, which optimizes a single detector for a constant background rate by maximizing $\sqrt{N}\frms$. Fig.~\ref{fig:strongly_modulated_unknown_multiple_detectors} shows that this orientation is very nearly optimal for even a strongly modulated background, provided that the three detectors are aligned with relative phases of approximately $\psi_{i+1} - \psi_{i} \approx 2\pi/3$.

\subsection{Sidereal Stacking} \label{sec:stackexample}

Over longer exposure times, the sidereal and solar days go out of phase with each other, and a single detector can distinguish the two types of signal without resorting to the 12-hour modulation of Fig.~\ref{fig:strongly_modulated_unknown_multiple_detectors}. 

\begin{figure}
\centering
\includegraphics[width=0.49\textwidth]{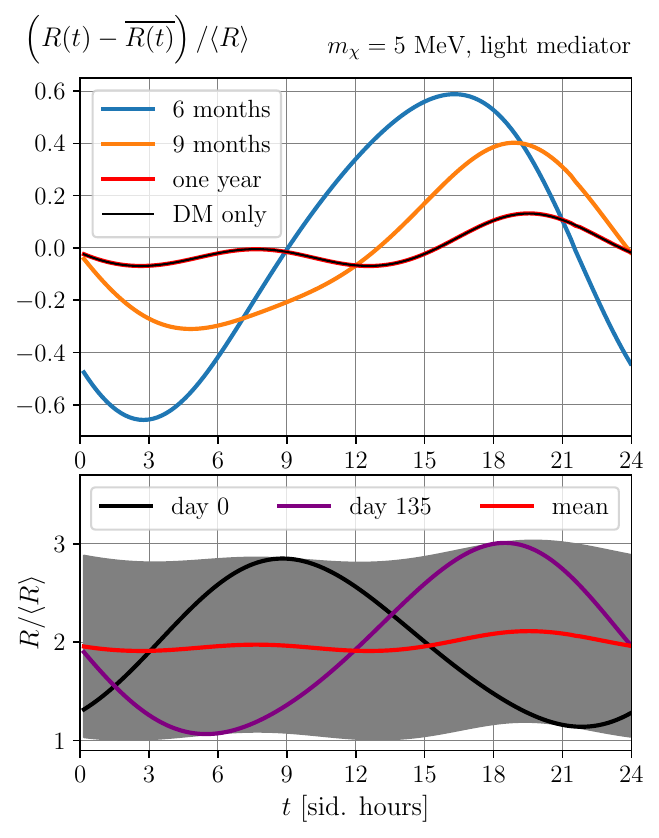}
\caption{ One year's worth of data (366 sidereal days) is stacked according to the sidereal time-of-day, in an example with a strong modulating background ($\frms(B) \approx 64\%$) with period 24.0\,h, and a smaller DM signal ($\frms(S) \approx 6.7\%$) modulating with the sidereal day. The black and purple lines in the lower panel show $R(t)$ on two particular days (Jan 1 and May 16), and the envelope of $R(t)$ for all other days of the year in gray. In red, we show the yearly average for $R(t)$, also as a function of sidereal time of day. In the upper panel, we show the stacked average for six months, nine months, or 366 sidereal days.  With a full year of data, the background cancels itself almost perfectly, so that the total rate matches the DM-only profile (shown in black). }
\label{fig:stacked}
\end{figure}

As an example, Fig.~\ref{fig:stacked} combines a DM signal with the strongly modulated SM background $B(t)$ of the previous section, given by \eqref{eq:bkgprofile} 
with $B_\text{osc} = 0.9 \, B_\text{flat}$ and an arbitrary phase. 

The $\frms$ score of this background model, $\frms(B) = 63.64\%$, is much larger than any of the DM signal models. Even so, a sufficiently long exposure time can allow us to extract the DM signal from this large, time-varying background. In this example, we take $B_\text{flat} = \langle R \rangle$ to be equal to the isotropic average DM rate for a 10\,MeV light mediator model, with a detector orientation that is approximately halfway between the best and worst orientations of Figure~\ref{fig:flatoptimal}. Averaged over the year, the DM signal $\frms$ is $6.67\%$, about nine times weaker than the background. 

The likelihood and test statistic in \eqref{eq:TSqA} and \eqref{eq:MultiplePoissonLikelihood} are valid for arbitrary exposure times and can be used without modification. To better visualize the data, however, we choose to stack the data over the sidereal time-of-day. 
As the lower panel of Fig.~\ref{fig:stacked} demonstrates, the combined signal-plus-background rate $R(t)$ is dominated by the background $B(t)$. However, over the course of a year the 24\,h modulation in $B(t)$ shifts out of phase with the sidereal day, destructively interfering with itself. With 1.0 years of data, the cancellation in the background is nearly exact: in the stacked analysis, the only impact from $B(t)$ is to add an effectively constant background of $B_\text{flat} \simeq \langle R \rangle$ to the DM signal.

For any other exposure time, the interference in the stacked $B(t)$ is no longer perfectly destructive, as the upper panel of Fig.~\ref{fig:stacked} shows. Even at nine months, the stacked signal is dominated by the modulating background. This result is specific to the strongly modulated case, where $\frms(B) > 9 \frms(S)$. For the DM signal to dominate the stacked rate, we need roughly 90\% of the background to cancel, i.e., the total exposure time should be more than 330 days. For shorter exposure times, we should simply perform the full \eqref{eq:fRMS_chi2} time-dependent analysis on the unstacked data.

\section{Conclusions}
\label{sec:Conclusions}

In this paper, we have constructed a general formalism for performing daily modulation analyses in directional dark matter detectors. Our analysis is sufficiently general to permit an unbinned analysis of time-series data for arbitrary numbers of detectors operated in parallel, with arbitrary flat or time-varying backgrounds either endogenous or exogenous to each detector. We have also presented several simplifications---for example, a binned analysis and a sidereal data-stacking procedure---which we hope will make our formalism practically useful for future directional detectors. The minimal ``recipe'' for computing an optimal sensitivity for multiple detectors is as follows:
\begin{itemize}
\item Given a binned time-dependent signal and background model (Eqs.~(\ref{eq:mu_mult_det}) and~(\ref{eq:SB_mult_det})), compute the Fisher information matrix elements in Eq.~(\ref{eq:Fisher_mult_det}).
\item Numerically maximize the right-hand side of Eq.~(\ref{eq:FisherVariance}) over detector configurations, which minimizes the variance of the signal-strength estimator, $\sigma_A^2$.
\item If the phase of the background is unknown, instead minimize the sensitivity marginalized over background phases, $\widetilde{\sigma_A}$, as in Eq.~(\ref{eq:sigmatilde}) 
\end{itemize}

In particular, we find that, based on assumptions about the character of the dominant background, even profiling over the phase of the background, a factor of $\sim 5$ in total exposure (either time, or detector mass if backgrounds scale with mass) can be achieved by optimizing the orientations of three detectors using the Fisher information. Note also that the configurations for weakly and strongly modulating backgrounds are highly similar, which is practically advantageous in that a configuration chosen with one background scenario in mind may be nearly optimal for many other modulating background scenarios. In our examples, the orientations that maximize Eq.~(\ref{eq:FisherVariance}) also very nearly maximize the $\frms$ of \eqref{eq:frms}, even for strongly modulated backgrounds. 

We have performed an example analysis for a fixed DM mass, but of course one may repeat the analysis using a prior on the distribution of DM masses, which may inform the best ``DM-weighted'' orientation of the experiment with which to take data. Furthermore, in a background-dominated experiment one could consider taking data for a short time (say, one day or one month), performing a likelihood analysis to determine the modulation phase under the assumption that the recorded counts are dominantly background, and subsequently taking data under the optimal orientation for the measured phase. This would likely lead to further improvements in the exposure required to claim a signal detection. However, in a multi-detector scenario, since the phase-agnostic expected sensitivity is comparable to the optimal sensitivity possible when the background phase is known perfectly, the sensitivity improvement may be somewhat marginal.

Not only can our detector optimization procedure handle large modulating backgrounds, it works even if the background oscillates at the same frequency as the sidereal day. With two or more directionally sensitive detector targets, rotated relative to each other, a well-designed experiment can break the degeneracy between the background and DM signal models. Of course, each experiment will have to contend with its own unknown backgrounds, but we hope that our methods are flexible enough to be adopted by directional detectors deployed in the coming years.

\section*{Acknowledgments}
We are grateful to the dark matter experimental community for enlightening conversations regarding daily modulation searches, especially (in alphabetical order) Dan Baxter, Miriam Diamond, Alex Drlica-Wagner, Nora Hoch, Edgar Marrufo Villalpando, Brandon Roach, Abigail Williams, and Lindley Winslow. We thank our chemistry colleagues Danna Freedman and Dane Johnson for numerous discussions regarding anisotropic organic scintillators. We thank Dan Baxter for reviewing an early version of this manuscript.
B.L.~thanks Ben Farr for helpful conversations. Y.K. acknowledges the support of a Discovery Grant from the Natural Sciences and Engineering Research Council of Canada (NSERC), as well as the Connaught New Researcher Award at University of Toronto. This work made use of computing resources provided by the Center for High Throughput Computing at the University of Wisconsin, Madison \cite{https://doi.org/10.21231/gnt1-hw21}.  This research used resources of the National Energy Research Scientific Computing Center (NERSC), a Department of Energy User Facility using NERSC award HEP-ERCAP 0037471. 

\bibliography{refs}

\end{document}